\documentclass[aps,prl,nobalancelastpage,superscriptaddress,twocolumn]{revtex4-2}
\usepackage{xr}
\usepackage[utf8]{inputenc}
\usepackage[english]{babel}
\usepackage[T1]{fontenc}
\usepackage[pdftex]{graphicx}  
\usepackage{xcolor}
\usepackage{physics}

\usepackage{tikz}
\usetikzlibrary{arrows.meta, bending}
\usepackage{dcolumn}
\usepackage{physics}

\usepackage{braket}
\usepackage{bm}
\usepackage{amsmath,amsthm,amssymb}
\usepackage{color}
\usepackage{verbatim}
\usepackage[normalem]{ulem}
\usepackage{dsfont}
\usepackage{ragged2e}
\usepackage{caption}
\usepackage{subcaption}
\DeclareCaptionJustification{forcejustify}{\leftskip=0pt \rightskip=0pt \parfillskip=0pt plus 1fill}
\usepackage{hyperref}
\newcommand{\be}{\begin{equation}}
\newcommand{\ee}{\end{equation}}
\hypersetup{
 colorlinks=true,
 linkcolor=blue,
 anchorcolor = blue,
 citecolor = blue,
 filecolor = blue,
 urlcolor = blue
}

\begin{document}
\title{
Space-time duality approach to (inhomogeneous) integrable quenches
}
\date{\today}
\author{Riccardo Travaglino}
\affiliation{SISSA and INFN Sezione di Trieste, via Bonomea 265, 34136 Trieste, Italy}

\author{Pasquale Calabrese}
\affiliation{SISSA and INFN Sezione di Trieste, via Bonomea 265, 34136 Trieste, Italy}

\author{Katja Klobas}
\affiliation{School of Physics and Astronomy, University of Birmingham, Edgbaston, Birmingham, B15 2TT, UK}

\author{Bruno Bertini}
\affiliation{School of Physics and Astronomy, University of Birmingham, Edgbaston, Birmingham, B15 2TT, UK}

\begin{abstract}

Characterising the universal aspects of non-equilibrium quantum many-body dynamics is one of the key goals of this century's physics research. Progress, however, is hindered by the lack of general theoretical frameworks for studying interacting quantum matter far from equilibrium. A recent breakthrough has been the realization that several key non-equilibrium quantities, such as the rate of growth of entanglement or the fluctuations of conserved charges within finite subsystems, can be related to equilibrium properties through a space–time duality that effectively exchanges the roles of space and time. This observation effectively enables the study of non-equilibrium phenomena using tools and concepts borrowed from equilibrium statistical mechanics and thermodynamics. A first proof of principle of this framework, dubbed `space-time duality approach' (SDA), was provided by interacting integrable systems, where thermodynamic properties can often be characterized exactly, while dynamical quantities typically remain beyond analytical reach. Subsequent developments, however, revealed that the SDA suffered from an intrinsic ambiguity, restricting its applicability to homogeneous quenches and to charge fluctuations arising from symmetric initial states. Here we resolve this ambiguity from first principles and derive closed-form predictions for entanglement growth and charge fluctuations after general quantum quenches. We benchmark our results against the exact analytical solution of the Rule 54 quantum cellular automaton and extensive TEBD simulations of the XXZ chain. Moreover we show that, when specialised to the entanglement entropy, our framework naturally reproduces the predictions of the quasiparticle picture.  

\end{abstract}
\maketitle

\emph{Introduction.---} The last quarter century has witnessed out-of-equilibrium quantum matter passing from being a mere theoretical abstraction to the subject of an active and rapidly expanding multidisciplinary research field. To date, some of the most important challenges in the field include understanding the principles governing the emergence of equilibrium from microscopic quantum dynamics~\cite{eisert2015quantum, calabrese2016introduction, serbyn2021quantum, bastianello2022introduction, abanin2019colloquium} and the dynamics of quantum information~\cite{calabrese2005evolution, nahum2017quantum, shenker2014multiple, kitaev2015talk, hosur2016chaos,calabrese_ln,zhou_nahum_2020}, characterising new forms of emergent universality~\cite{skinner2019measurement, li2019measurement} and novel out-of-equilibrium phases of matter~\cite{sacha2018time, khemani2019brief}, and achieving fault-tolerant quantum computation~\cite{preskill2018quantum, altman2023quantum}. 

From a theoretical perspective, the main obstacle to addressing these questions is the lack of general methods for describing interacting quantum many-body systems out of equilibrium. In generic spatial dimension, numerical approaches based on classical computation face unavoidable fundamental limitations~\cite{rigol2008thermalization, carleo2012localization, James_2015,feynman2018simulating,Murciano_2022,Lin22,Krinitsin25,Tindall_2026}, while existing (semi)analytical techniques typically rely on severe and often uncontrolled approximations~\cite{berges2004prethermalization,Sotiriadis_2010,Schiro10,Gambassi_2011,Chiocchetta15,aoki2014nonequilibrium, dalfovo1999theory}. Even in one dimension, where advanced tensor-network methods can access remarkably large systems~\cite{vidal2003_tebd,Daley_2004,Schollwock_2011, white2004real,Hauschild_2018,iTensor}, the simulation of quantum dynamics remains restricted to relatively short times: matrix-product-state methods are limited by the growth of entanglement entropy, while matrix-product-operator approaches face the  infamous entanglement barrier~\cite{dubail2017entanglement,Rath_2023}.

A recent breakthrough came with the realisation that some relevant non-equilibrium properties can be described efficiently by exchanging the roles of space and time. The key observation is that, upon interchanging space and time, the evaluation of finite-time properties in large systems can be mapped onto the study of stationary properties in finite systems~\cite{bertini2022growth,bertini2025exactly}.  The price to pay is that the original unitary time-evolution operator is replaced by a generally non-unitary transfer matrix. Although this approach was originally introduced as a numerical technique~\cite{banuls2009matrix,Muller_2012}, it was subsequently recognised as a powerful framework for investigating general aspects of quantum many-body dynamics~\cite{ippoliti2021fractal,lerose2021influence,bertini2022entanglement} and, in some cases, for obtaining exact analytical results~\cite{bertini2018exact,bertini2019exact,bertini2025exactly}. 

\begin{figure*}[t]
  \includegraphics[width=\linewidth]{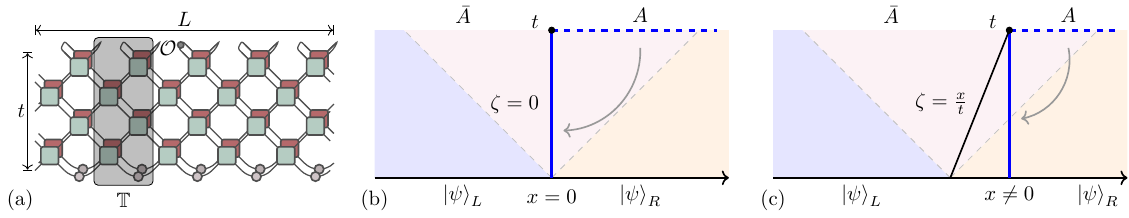}
  \caption{\label{fig:zetaneq0} (a): A tensor-network diagram of a local expectation value after a quantum quench under circuit evolution. Green and red gates represent $U$ and $U^{\ast}$ respectively, gray circles at the bottom are the initial (product) state, and the circle at the top is a local observable $\mathcal{O}$. Equivalently the tensor network can be contracted \emph{in space}  by replacing the space-transfer matrix $\mathbb{T}$ with the projector to its leading eigenvectors. (b): Duality from the original entropy of $A$ at time t (dashed blue line) to the entropy in the dual channel at the boundary of $A$ (full blue line).  The light cone represents the region in which the occupation functions are a mixture of those of left and right states. Since the boundary between $A$ and $\bar{A}$ lies precisely at the bipartition, in the swapped channel the solution depends solely on the ray $\zeta=0$. (c): Duality from the original entropy of $A$ at time t to the entropy in the dual channel at the boundary of $A$ for $x\neq0$. Since the boundary between $A$ and $\bar{A}$ is distinct from the bipartition point, the theory in swapped channel involves all rays in $[{x}/{t}, \infty]$. }
\end{figure*}

A particularly fruitful arena where these ideas found application is that of interacting integrable systems, where one can describe stationary properties through exact analytical formulas~\cite{yangyang,takahashi1999book,Caux_2013,Ilievski_2015,Caux_2016,alba_quench_action_2017}, while dynamical properties are generically out of reach for both analytical and numerical approaches (with the exception of the late-time, `hydrodynamic' regime~\cite{castroalvaredo2016emergent,bertini2016transport,doyon2020lecture,Alba_2021,Essler_2023} and the quasiparticle picture for entanglement entropy \cite{calabrese2005evolution,alba_entanglement_2017,calabrese_ln,alba2}). In these systems the exchange of space and time, dubbed `space-time duality approach' (SDA), provided new closed form predictions for quantities that were considered beyond the capacity of analytical methods such as the rate of growth of R\'enyi entanglement entropies~\cite{bertini2022growth} and the behaviour of charge fluctuations~\cite{bertini2023nonequilibrium} after quantum quenches from integrable initial states~\cite{piroli2017what}. Further research, however, revealed that the SDA suffered from an ambiguity that limited its applicability to homogeneous quenches and charge fluctuations in homogeneous initial states~\cite{bertini2024dynamics,horvath2024full,horvath2025counting,horvath2025a}. In essence, this ambiguity was caused by the fact that in the `dual system' the momentum of quasiparticles is no longer a monotonic function of the rapidity. 

Here we resolve this ambiguity by developing an \emph{ab-initio} thermodynamic Bethe ansatz (TBA) treatment of systems with non-monotonic quasiparticle momentum. Building on this framework, we derive closed-form analytical expressions describing both entanglement growth and the dynamics of charge fluctuations following general inhomogeneous quenches in integrable systems. We argue that these results are \emph{exact}: they reproduce all known predictions in the non-interacting limit~\cite{bertini2018entanglement,bertini2024dynamics}, as well as in the interacting quantum cellular automaton Rule 54~\cite{klobas2021exact,klobas2021entanglement,klobas2024non}, while displaying excellent agreement with TEBD simulations in the XXZ spin chain. Furthermore, we show that only after resolving the ambiguity according to our prescription do the SDA expressions for R\'enyi entropies recover, in the von Neumann limit, the predictions of the quasiparticle picture~\cite{alba_entanglement_2017,alba2019entanglement}. At the same time, for any R\'enyi index different from one, the resulting expressions remain genuinely beyond the scope of the quasiparticle framework, mirroring what was previously observed in the homogeneous setting~\cite{bertini2022growth}.

\emph{Setting.---} We consider a one-dimensional quantum system prepared in a spatially inhomogeneous non-equilibrium state $\ket{\psi_0}$ and then let to evolve according to the integrable Hamiltonian $H$. The initial state is assumed to have short range correlations, and to be integrable according to Ref.~\cite{piroli2017what}. For example, one can think of a `bipartitioning protocol' producing an initial state of the form $\ket{\psi_0} = \ket{\psi_{\rm L}} \otimes\ket{\psi_{\rm R}}$, where $\ket{\psi_{\rm L}}$ and $\ket{\psi_{\rm R}}$ are homogeneous states on the left and right halves of the system, see Fig.~\ref{fig:zetaneq0}. The homogeneous limit is obtained by setting $\ket{\psi_{\rm L}} = \ket{\psi_{\rm R}}$.

Integrability ensures that the Hamiltonian’s spectrum is constructed in terms of single-particle excitations~\cite{korepin1993book} and the equilibrium properties of the system are described by the TBA~\cite{yangyang,Zamolodchikov1989,takahashi1999book}. In contrast with noninteracting systems, however, the single particle excitations in a given many-body state experience non-trivial two-body interactions. Specifically, in the TBA framework one describes eigenstates $\ket{\{\rho_{\rm t},\vartheta\}}$ using two functions: the `total root density' $\rho_{\rm t}(\mu)$, and the `filling function' $\vartheta(\mu)$ which are related through a set of integral equations which account for the interactions~\cite{Note50}. The first represents the density of single particle states and the second the portion of these states that are occupied. The rapidity variable $\mu$ represents a convenient parameterisation of the single particle dispersion relation $(p(\mu),\varepsilon(\mu))$~\cite{orbach1958}. It is also useful to introduce the `dressed velocity' $v^{\rm dr}(\mu)$ representing the velocity of an elementary particle added to the state $\ket{\{\rho_{\rm t},\vartheta\}}$~\cite{bonnes2014}. Because of the interactions, $v^{\rm dr}(\mu)$ differs from the `bare velocity' $v(\mu)=\varepsilon'(\mu)/p'(\mu)$ of the quasiparticle and it is related to the latter via an integral equation involving $\rho_{\rm t}(\mu)$ and $\vartheta(\mu)$~\cite{Note50}. 

The Generalised Hydrodynamics (GHD) framework~\cite{castroalvaredo2016emergent,bertini2016transport,doyon2020lecture} allows one to extend this approach to describe the stationary states reached at late times by local subsystems after inhomogeneous quenches. The quasi-stationary state at the space-time point $(x,t)$ is specified by means of the space-time dependent total root densities and filling functions, $\{\rho_{{\rm t}}(\mu,x,t),\vartheta(\mu,x,t)\}$, fulfilling (non-linear) continuity equations involving $v^{\rm dr}(\mu)$~\cite{Note50}. In the particular case of a bipartitioning protocol $\{\rho_{{\rm t}},\vartheta\}$ depend on $(x,t)$ only through the space-time ray $\zeta =x/t$, where $x$ is the distance from the junction. 

Importantly, despite the fact that they explicitly depend on time, the functions $\{\rho_{{\rm t}}(\mu,x,t),\vartheta(\mu,x,t)\}$ only describe the state of the system at asymptotically large times, when local subsystems have already reached equilibrium. In this regime the residual time dependence is only due to the inhomogeneity of the system and is accounted for by classical hydrodynamic equations. This `equilibrium' regime is entirely distinct from the genuine `non-equilibrium' regime that we aim to study in this Letter and a priori retains no information on the growth of entanglement or the evolution of correlation functions. Describing the non-equilibrium regime in the presence of interactions is currently an open problem in integrable systems: here we argue that the SDA solves this problem in the case of integrable initial states.

\emph{Space-Time Duality.---}  The main premise of the SDA is that the non-equilibrium regime can be mapped to the stationary regime of a dual theory in which the role of space and time are exchanged. Whenever the latter is integrable (hence the requirement of an integrable initial state) one can then use TBA and GHD to characterise this equilibrium regime and therefore the non-equilibrium regime of the original theory. 

The origin of this duality is better understood in terms of models with discrete time dynamics, where space and time are treated on similar footings and at the same time avoiding the complications of continuum theories~\cite{bertini2022growth, bertini2023nonequilibrium}. Consider for instance the one-point function of a one-site operator $\mathcal O$ on the state at time $t$, which is represented diagrammatically as in Fig.~\ref{fig:zetaneq0}(a). As it is clear from the diagram, this object can be computed by contracting the tensor network horizontally. Namely, one can express it using a suitably defined \emph{space transfer matrix} $\mathbb{T}$, i.e.\ a super-operator acting on the (doubled) temporal lattice. In particular, in the thermodynamic limit we find 
\begin{equation} \label{eq:EV}
\lim_{L\to\infty}\expval{\mathcal O}{\Psi_t} = {\rm tr}[\tilde \rho_{{\rm st}, t} \mathcal O].
\end{equation}
Here we set $\tilde \rho_{{\rm st}, t}=R L$, where $R$ and $L$ are the (left and right) leading eigenoperators of the space transfer matrix (which are unique under mild conditions on the initial state~\cite{klobas2021exact2}) and $t$ is the length of the lattice in which $\tilde \rho_{{\rm st}, t}$ is defined. Similarly, considering the half system reduced density matrix, one can show that its R\'enyi entropies fulfil (for open boundary conditions)~\cite{bertini2022growth}
\begin{equation}
\label{eq:tracepowers}
S_\alpha(t)=
  \frac{\log\tr[  \rho_{[0,\infty]}(t)^\alpha]}{1-\alpha} 
  =\frac{\log \tr[\tilde \rho_{{\rm st}, t}^\alpha]}{1-\alpha},
  %\,\,\, \forall \alpha\in \mathbb N, 
\end{equation}
where we explicitly took the thermodynamic limit. Finally, if the original system has a $U(1)$ charge $Q$, we also have the full counting statistics of the current at $x=0$ is given by~\cite{bertini2023nonequilibrium,bertini2024dynamics} (for open boundary conditions)
\begin{equation} \label{eq:FCS}
Z_\beta(t)\mkern-4mu=\mkern-8mu
  \mel{\Psi_0}{e^{\beta Q_{[0,\infty]}\!(t)}e^{-\beta Q_{[0,\infty]}}}{\Psi_0}
  \mkern-6mu=\mkern-4mu 
  \tr\mkern-4mu
  \big[\tilde \rho_{{\rm st}, t}  e^{\beta \tilde Q_{[0,t]}}\mkern-2mu\big],
\end{equation}
where $\tilde Q$ is the charge of the dual system and the subscript denotes the interval over which the charge is computed. Eqs.~(\ref{eq:EV}--\ref{eq:FCS}) provide a complete description of expectation values, R\'enyi entropies, and charge fluctuations in the non equilibrium regime. The analogous statements for continuous time follow by taking a Trotter limit (i.e.\ continuum limit for the time lattice). Note that in models with continuous space-time, similar features can be argued using the topological nature of the line connecting the twist fields (or charge lines)~\cite{doyon2025twistfieldsmanybodyphysics}.

Interpreting  $\tilde \rho_{{\rm st}, t}$ as a stationary state one then conjectures the following correspondence 
 \be
 \tr[\tilde \rho_{{\rm st}, t}^\alpha \mathcal O] \overset{ t \leftrightarrow x}{\longleftrightarrow} \tr[\rho_{{\rm st}, x}^\alpha  \mathcal O], 
 \label{eq:swap}
 \ee
where $\mathcal O$ is an arbitrary observable and $\rho_{{\rm st}, x}$ is the ``regular'' stationary state of the time evolution reduced to a subsystem of length $x$ and $\overset{ t \leftrightarrow x}{\longleftrightarrow}$ indicates the swap of space and time. Note that this swap bears some conceptual similarities with the mirror approaches used in relativistic quantum field theory and string theory~\cite{Zamolodchikov1989,Bombardelli_2009,Tongeren_2016}, although, contrary to the latter approaches, the exchange is performed in real rather than imaginary time. 

In practice, Eq.~\eqref{eq:swap} comprises two main steps~\cite{bertini2022growth}. Firstly, one assumes that the filling function $\vartheta(x,t)$ has the same functional form in the original and swapped theories. Secondly, one performs the exchange of space and time through a swap of the dispersion relation, i.e.
\begin{equation} \label{eq:PEswap}
p(\mu) \leftrightarrow \varepsilon(\mu). 
\end{equation}
This simple mapping, however, leads to a subtle ambiguity: to have a well-defined total root density $\rho_{\rm t}(\mu)$ the standard TBA treatment requires a strictly monotonic single-particle momentum $p'(\mu)>0$, therefore the equations are blind to ${\rm sgn}(p'(\mu))$. Since the single particle energy is generically not monotonic the TBA equations for the dual theory are not uniquely defined by the swap in Eq.~\eqref{eq:PEswap}, as they can involve arbitrary insertions of ${\rm sgn}(\varepsilon'(\mu))$ or of ${\rm sgn}(v^{\rm dr}(\mu))$~\cite{bertini2024dynamics}. This ambiguity was reduced in Ref.~\cite{bertini2022growth} by requiring the SDA predictions to reproduce the exact results of Ref.~\cite{klobas2021entanglement} for the dynamics of a special interacting integrable quantum cellular automaton. This requirement, however, does not distinguish between ${\rm sgn}(\varepsilon'(\mu))$ and  ${\rm sgn}(v^{\rm dr}(\mu))$ and, as a result, it only resolves the ambiguity when the latter are the same, i.e.\ for homogeneous integrable quenches~\cite{bertini2024dynamics, horvath2024full, horvath2025counting}. 

This issue can be resolved by deriving TBA expressions for systems with non-monotonic momentum.
%Here we completely resolve this issue by deriving TBA expressions for systems with non-monotonic momentum. %Instead of sub-dividing the rapidity interval in sub-intervals where $p(\mu)$ is strictly monotonic and treating each of such branches as a different quasiparticle species as in the standard approach~\cite{takahashi1999book}, here 
To this end, we perform a suitable rapidity re-parametrisation $\mu \to f^{-1}(\mu)=\phi$, such that $\tilde p(\phi) \equiv p(f(\phi))$ is monotonic (see the End Matter for details). This gives a single set of TBA equations involving factors of $\text{sgn} \,p'(\mu)$ to account for the non-monotonicity of the momentum. Computing the relevant quantities and performing the swap in Eq.~\eqref{eq:PEswap} this gives the following prediction for the asymptotic slope of R\'enyi entropies
\begin{equation} \label{eq:slopeSDA}
  \mkern-8mu
\begin{aligned}
  s_\alpha &= \lim_{t\to\infty} \frac{S_\alpha(t)}{t} =\frac{1}{1-\alpha}\int \frac{{\rm d}\mu}{2\pi} \varepsilon'(\mu)\mathcal{L}_\alpha(\mu),
   \\
    \mathcal{L}_\alpha(\mu) &= \text{sgn}(\varepsilon'(\mu))
    \log\biggl[(1\!-\!\vartheta(\mu)) ^\alpha \mkern-6mu+\mkern-4mu 
    \frac{\vartheta(\mu)^\alpha}{y_\alpha(\mu)^{\text{sgn}(\varepsilon'(\mu))}}\biggr],
    \end{aligned}
  \mkern-8mu
\end{equation}
where the auxiliary function $ y_\alpha $ is defined self consistently through the integral equation
\begin{equation}
\label{eq:yEq}
  \log y_\alpha(\mu) = \int \frac{{\rm d}\mu'}{2\pi}K(\mu',\mu)\mathcal{L}_\alpha(\mu'),
\end{equation}
where $K(\nu,\mu)$ is the scattering kernel of the system (logarithmic derivative of the scattering matrix). Analogously, we find that for large times the full counting statistics takes a large deviation form
\begin{equation} \label{eq:largedeviationFCS}
Z_\beta(t) \simeq e^{t z_\beta}, 
\end{equation}
with exponent given by 
\begin{equation}
  z_\beta=\lim_{t\to\infty} \frac{1}{t} \log {Z_\beta(t)} =  \int \frac{{\rm d}\mu} {2\pi} \varepsilon'(\mu) \mathcal{L}^{(\beta)}_1(\mu), 
  \label{eq:FCSSDA}
\end{equation}
where $\mathcal{L}^{(\beta)}_1$ takes the same form as $\mathcal{L}_\alpha(\mu)$ but with a modified auxiliary function to account for the charge insertion
\begin{equation}
     \log y_{1}^{(\beta)}(\mu) = -\beta \tilde q(\mu) + \int {\rm d}\mu' K(\mu',\mu)\mathcal{L}_1^{(\beta)}(\mu') .
\end{equation}
Equation~\eqref{eq:FCSSDA} differs from the prediction of ballistic fluctuation theory (BFT) for the full counting statistics on equilibrium states~\cite{myers2020}. The difference, however, is very subtle and comes from using in $\mathcal{L}^{(\beta)}_1$ the sign of ${\rm sgn}(\varepsilon'(\mu))$, i.e.\ of the bare velocity, instead of that of the dressed velocity (in the state with filling function $\vartheta(\mu)/((1-\vartheta(\mu))y_{1}^{(\beta)}(\mu)^{v^{\rm dr}(\mu)}+\vartheta(\mu))$~\cite{bertini2024dynamics}). Note that the BFT expression was argued not to hold after a quench whenever ${\rm sgn}(\varepsilon'(\mu)) \neq {\rm sgn}(v^{\rm dr}(\mu))$, i.e.\ for non-symmetric or inhomogeneous initial states, because quasiparticle trajectories are generically curved~\cite{horvath2025counting}. Instead, Eq.~\eqref{eq:slopeSDA} is precisely the expression postulated in Ref.~\cite{bertini2022growth}: our results imply that the latter should hold also when ${\rm sgn}(\varepsilon'(\mu)) \neq {\rm sgn}(v^{\rm dr}(\mu))$.  

To test our formulas, we then focus on the case where ${\rm sgn}(\varepsilon'(\mu))$ and ${\rm sgn}(v^{\rm dr}(\mu))$ are maximally different, i.e.\ that of inhomogeneous quenches.
As shown in the supplemental material (SM) one can describe inhomogeneous quenches by integrating Eqs.~\eqref{eq:slopeSDA} and \eqref{eq:FCSSDA} over each separate fluid cell (each one with filling function $\vartheta(\mu, x,t)$)~\cite{Note50}. The general result can also be specialised to the simpler case of the bipartitioning protocol, which allows for a direct numerical evaluation. In particular, denoting by $-x$ the position of the bipartition point between $\ket{\psi_L}$ and $\ket{\psi_R}$ (cf.~Fig.~\ref{fig:zetaneq0}) we find that the entropy scales as
\begin{equation}
S_{[x,\infty]}^{(\alpha)} (t) = S_{[\zeta t,\infty]}^{(\alpha)} (t) \sim t s_{\alpha}(\zeta), 
\end{equation}
where we introduced the `ray' $\zeta=x/t$. 

In the case $x=0$ the result turns out to only depend on the ray $\zeta=0$ (see~Fig.~\ref{fig:zetaneq0}). Namely, we have that $s_{\alpha}(\zeta=0)$ is described by Eq.~\eqref{eq:slopeSDA} with $\vartheta(\mu)\mapsto\vartheta(\mu,\zeta=0)$. Instead, the solution for different values of $x$ involves the GHD solution at different rays (see Fig.~\ref{fig:zetaneq0}). By means of simple algebraic manipulations the result can be expressed as~\cite{Note50}
\begin{equation}
 s_{\alpha}(\zeta) = \frac{1}{1-\alpha}\int \frac{{\rm d}\mu}{2\pi} (\varepsilon'(\mu)-\zeta p'(\mu)) \mathcal{L}_\alpha(\mu,\zeta),
\label{eq:Sgeneralzeta}
\end{equation}
where $\mathcal{L}_\alpha(\mu,\zeta)$ is obtained from the one in Eq.~\eqref{eq:slopeSDA} with the replacements $\vartheta(\mu)\mapsto \vartheta(\mu,\zeta)$ and $y_\alpha(\mu)\mapsto y_\alpha(\mu, \zeta)$~(that is found by solving Eq.~\eqref{eq:yEq} with $\mathcal{L}_\alpha(\mu)\mapsto \mathcal{L}_\alpha(\mu,\zeta)$). Note that Eq.~\eqref{eq:Sgeneralzeta} is almost a transformation to the co-moving frame of the ray $\zeta$, i.e.\ $\varepsilon'(\mu)\mapsto \varepsilon'(\mu)-\zeta p'(\mu)$, but not exactly as $\mathcal{L}_\alpha(\mu,\zeta)$ still involves ${\rm sgn}(\varepsilon'(\mu))$. Analogously, we have that the full counting statistics retains the form in Eq.~\eqref{eq:largedeviationFCS} with $z_\beta$ replaced by 
\be
z_\beta(\zeta) = \int \frac{{\rm d}\mu}{2\pi} (\varepsilon'(\mu)-\zeta p'(\mu))  \mathcal{L}^{(\beta)}_1(\mu,\zeta)\,. 
\label{eq:FCSgeneralzeta}
 \ee

%\emph{Analytical Tests.---}
\emph{Consistency Checks.---}
We begin by considering some relevant limits of Eqs.~\eqref{eq:Sgeneralzeta} and \eqref{eq:FCSgeneralzeta}. First we note that if $|x|\geq v_{\rm max} t$, where $v_{\rm max}$ is the maximal velocity of the quasiparticles, the equations should reduce to the homogeneous case as a consequence of the Lieb-Robinson bound. This is indeed the case: for $|\zeta| > v_{\rm max}$ one has that $\vartheta(\mu,\zeta)$ is either equal to the filling function of $\ket{\psi_{\rm L}}$ or of $\ket{\psi_{\rm R}}$, implying that it is a symmetric function of the rapidity (both $\ket{\psi_{\rm L}}$ and $\ket{\psi_{\rm R}}$ are integrable states~\cite{piroli2017what}). Therefore the term in $\zeta$ vanishes by antisymmetry of the integrand and one recovers the homogeneous result. For the same reason, one obtains the homogeneous result for all $\zeta$s in the special case $\ket{\psi_L} = \ket{\psi_R}$. Next, we note that Eqs.~\eqref{eq:Sgeneralzeta} and \eqref{eq:FCSgeneralzeta} reproduce the exact results obtained for inhomogeneous quenches in both free fermions~\cite{bertini2018entanglement, bertini2024dynamics} and the (interacting) quantum cellular automaton Rule 54~\cite{klobas2021entanglement, klobas2024non}. Finally, considering $\alpha\to 1$ we find that Eq.~\eqref{eq:Sgeneralzeta} reduces to the quasi-particle picture result for the von-Neumann entropy in interacting integrable models~\cite{alba2019entanglement}. This is a particularly important test as the result of Ref.~\cite{alba2019entanglement} accounts for curved quasiparticle trajectories, i.e\ the fact that ${\rm sgn}(\varepsilon'(\mu)) \neq {\rm sgn}(v^{\rm dr}(\mu))$. 

%\emph{Numerical Tests.---}
To test our results numerically we consider the paradigmatic example of the XXZ chain 
\begin{equation}
    H_{\rm XXZ} = \sum_j \sigma_j^x\sigma_{j+1}^x + \sigma_j^y\sigma_{j+1}^y +\Delta \sigma_j^z\sigma_{j+1}^z,
\end{equation}
where $\sigma_j^{x,y,z}$ act as Pauli matrices at site $j$ and the anisotropy $\Delta$ is a real parameter, and study bipartitioning protocols joining states of the (tilted) N\'eel and (tilted) ferromagnetic families
\begin{equation}
  \mkern-8mu
  \ket{N,\theta} = e^{i\frac{\theta}{2}\sum_j \sigma_j^x} 
  \ket{\uparrow\downarrow}^{\otimes \frac{L}{2}}\mkern-4mu,\mkern12mu
  %\ket{\uparrow\downarrow\uparrow\downarrow \dots},&\quad
 \ket{F,\theta} =  e^{i\frac{\theta}{2}\sum_j \sigma_j^x}
  \ket{\uparrow}^{\otimes L}\mkern-8mu.
  %\ket{\uparrow\uparrow\uparrow\uparrow\dots}.
  \mkern-8mu
\end{equation}
These states are integrable and, for $\theta\neq 0, \pi$, not symmetric under the $U(1)$ charge $\sum_j \sigma_j^z$ which characterises the XXZ chain. 
A third class of initial states which we consider is the Majumdar-Ghosh (dimer) state,
\begin{equation}
     \mkern-8mu
  \ket{D} = \frac{1}{2^{\frac{L}{2}}}
  \Big(\ket{\uparrow\downarrow} - \ket{\downarrow\uparrow}\Big)^{\otimes \frac{L}{2}}\mkern-4mu,\mkern12mu 
\end{equation}
which is integrable and symmetric. 
All TBA results for homogenous quenches can be found in \cite{piroli2016,Bertini_2018_lowtransport}. Furthermore, in the SM we also consider bipartitions of thermal states at different temperatures, a setup which is commonly considered to study transport~\cite{castroalvaredo2016emergent,bertini2016transport}.

The time evolution is implemented via time-evolving block decimation (TEBD) algorithm~\cite{vidal2003_tebd,Daley_2004,Schollwock_2011}, which allows us to reach times of about $t\sim 15$ in most quenches of interest by describing the states as matrix product states (MPS) with maximum bond dimension $\chi_{\rm max}=1024$.  We are interested in the comparison between the results of this numerical evolution and those of Eqs.~\eqref{eq:Sgeneralzeta} and \eqref{eq:FCSgeneralzeta}. For reference we also report the results obtained by replacing ${\rm sgn}(\varepsilon'(\mu))$ with ${\rm sgn}(v^{\rm dr}(\mu))$, i.e.\ deliberately missing the effect of curved quasiparticle trajectories. Since the initial time regimes exhibit strong oscillations which strongly affects the global slopes in Eqs.~\eqref{eq:slopeSDA} and  \eqref{eq:FCSSDA} within the time regimes accessible to TEBD, we numerically evaluate the instantaneous slopes ${{\rm d} S^{(\alpha)}}/{{\rm d} t}$ and ${{\rm d}\log Z_\beta}/{{\rm d}t}$. The two approaches give the same asymptotic behaviour because of the linear growth.

\begin{figure}[h!]
    \centering
    \includegraphics[width=\linewidth]{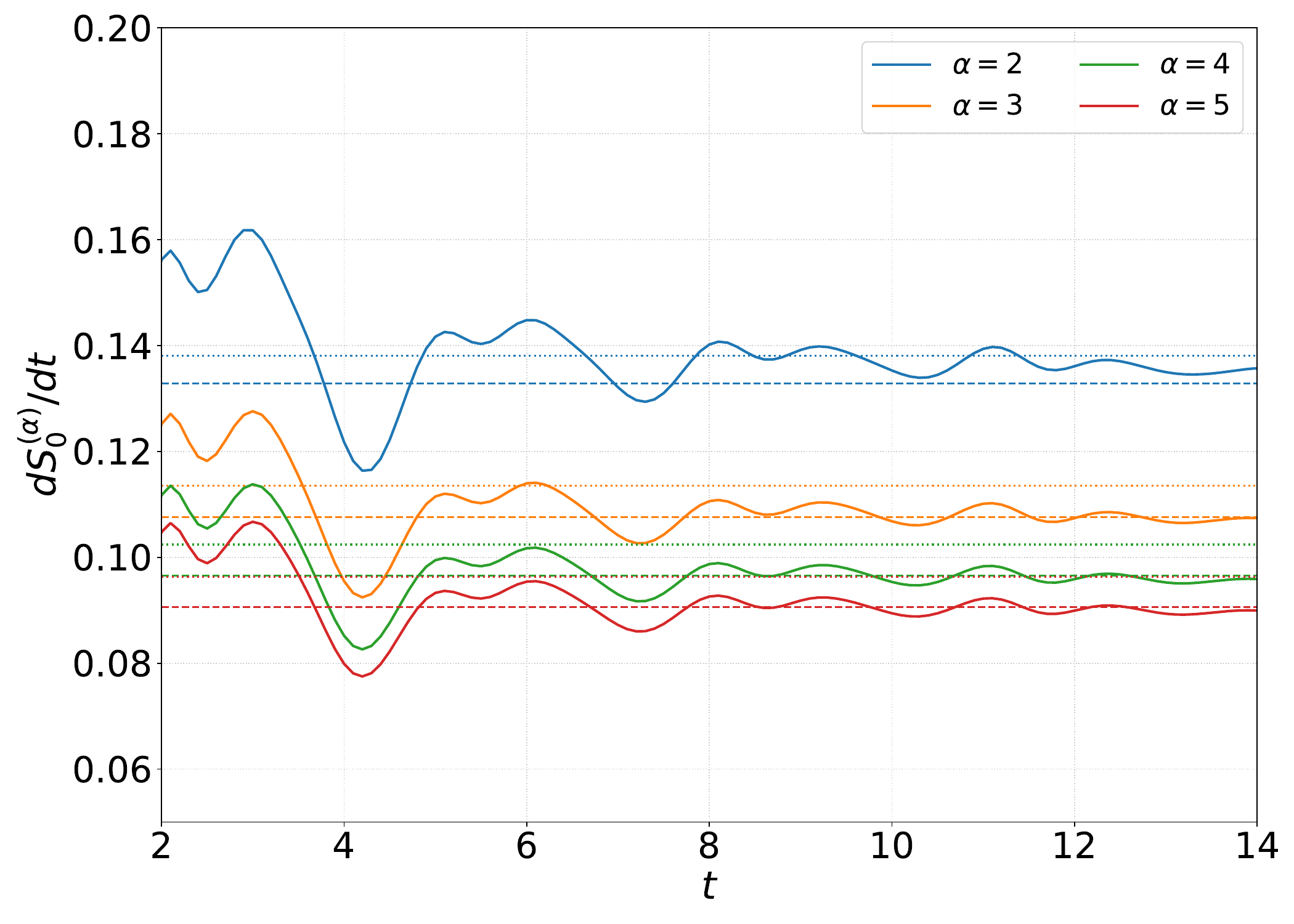}
    \caption{Instantaneous slope of the R\'enyi entropies of $[0,\infty]$ after a quench from the bipartite state $\ket{N,0}_L\otimes\ket{N,{\pi}/{12}}_R$ in the XXZ chain with $\Delta=4$. Dashed lines represent the predictions obtained by using $\text{sgn}(\varepsilon'(\theta))$, while dotted lines are the predictions obtained using $\text{sgn}(v^{\rm dr}(\theta))$.}
    \label{fig:comparisonpi12}
\end{figure}

Considering first the case of R\'enyi entropies (see a representative example in Fig.~\ref{fig:comparisonpi12} and further ones in the SM) we see almost perfect agreement already at short times, which is surprising as Eq.~\eqref{eq:Sgeneralzeta} is an asymptotic prediction. Moreover, although for $\alpha=1,2$ the numerics are not able to distinguish between the predictions using ${\rm sgn}(\varepsilon'(\mu))$ and  ${\rm sgn}(v^{\rm dr}(\mu))$, they clearly agree only with the former for $\alpha\geq 3$, confirming the validity of Eq.~\eqref{eq:Sgeneralzeta} also in the non-trivial case where quasiparticle trajectories are curved. Note that the Rényi entropies with $\alpha\geq 3$ are less dependent on the smaller Schmidt values, which are truncated in the TEBD approach, and give therefore more reliable results.

\begin{figure}
    \centering
    \includegraphics[width=\linewidth]{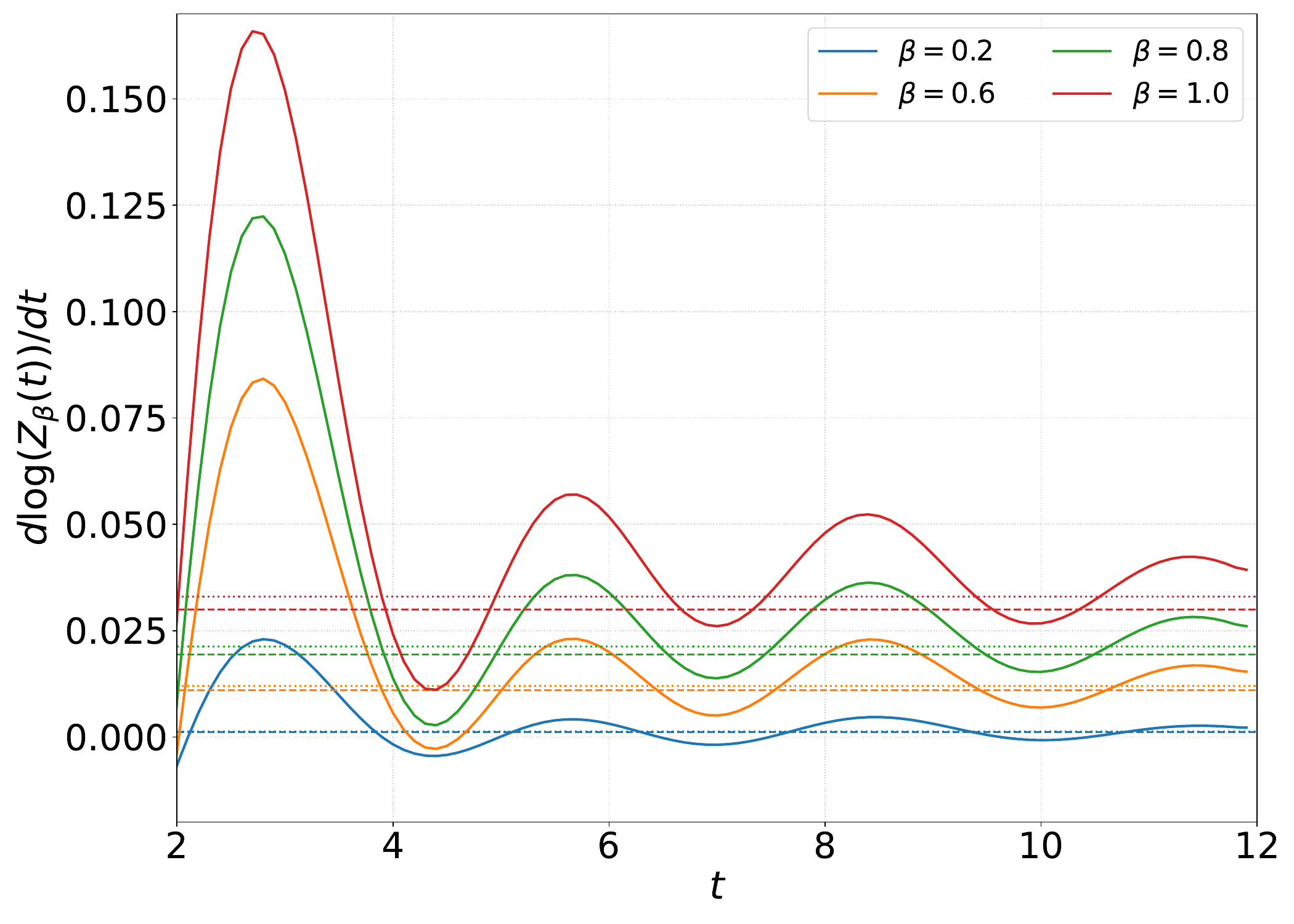}
    \caption{Instantanoeus slope of the FCS of the left half $[-\infty,0]$ in a quench from the state $\ket{N,0}_L\otimes\ket{D}_R$ in the XXZ chain with $\Delta =0.5$. While the numerics exhibits good convergence to the asymptotic values predicted, strong oscillations make it impossible to distinguish between the two sign choices on the time intervals accessible via TEBD.}
    \label{fig:comparisonneeldimer}
\end{figure}

On the other hand, the numerics is less conclusive for the FCS, as seen in Fig.~\ref{fig:comparisonneeldimer} (additional examples of quenches from both pure and thermal bipartitioned states are reported in the SM). While the numerical data clearly converges to the asymptotic results, strong oscillations make it difficult to distinguish between the two predictions, whose difference is smaller than the oscillations' amplitude. Similar numerical issues were noted in \cite{bertini2024dynamics} for the global slope in a homogeneous quench, suggesting that such strong early-time oscillations are a general feature of FCS evaluations.

\emph{Conclusions.---} In this letter we have resolved from first principles an ambiguity affecting the space-time duality approach to integrable quenches~\cite{bertini2022growth, bertini2023nonequilibrium, bertini2024dynamics} and used it to obtain closed-form predictions for the large-scale entanglement growth and the evolution of charge fluctuations in the presence of both non-trivial interactions and spatial inhomogeneity. We have shown that our predictions recover known exact results, whenever the latter are available, and agree very well with extensive TEBD simulations of the dynamics of the XXZ spin-$1/2$ chain. Therefore, they seem to provide a hitherto missing exact description of interacting integrable dynamics in the presence of interactions. 

Our results prompt two immediate questions for future research. The first is to provide a mathematical proof of our predictions by developing an exact characterisation of the thermodynamic properties the fixed points of the space evolution. The second is whether our approach can be generalised to other relevant probes of the non-equilibrium dynamics which are currently beyond reach, e.g.\ equal time two-point functions.

\begin{acknowledgments}
\emph{Acknowledgments.---}
We thank Benjamin Doyon, David Horvath, Samuel Pickering and Colin Rylands for useful discussions.  We acknowledge financial support from the Royal Society through the University Research Fellowship No.\ 201101 (B.B.), and the European Commission through the ERC-AdG grant MOSE No.\ 101199196 (R.T.\ and P.C.). R.T.\ thanks the physics department of the University of Birmingham for the hospitality when the core of this work was carried out.
\end{acknowledgments}
\begin{center}
\begin{large}
\textbf{End Matter}
\end{large}
\end{center}

\emph{TBA with Non-monotonic Momentum.---}
In this section we obtain a generalisation of TBA equations for a model with non-monotonic momentum. For simplicity, we consider a model with a single species of quasiparticles, as the modifications accounting for multiple species are straightforward. 

Within the TBA description thermodynamics is constructed by placing the theory on a ring with periodic boundary conditions, which impose the quantisation condition for the momenta
\begin{equation}
  z(\mu_i) \equiv \frac{1}{2\pi} p(\mu_i) +(\delta*\rho_p)(\mu_i) = \frac{n_i}{L}, 
\end{equation}
where $\delta$ is the scattering phase, $n_i$ are integer numbers which specify the state, while $\mu_i$ represents the rapidity variable in the parametrisation which is more convenient for each specific model, and $\rho_p$ is the TBA particle density. 

Typically one considers rapidity variables defined such that the condition $p'(\mu)>0$ is satisfied and this can be used to define the density of states $\rho_{\rm t}$ in the thermodynamic limit 
\begin{equation}
    \frac{{\rm d}z(\mu)}{{\rm d}\mu} = \frac{1}{2\pi} p'(\mu) + (K*\rho_p)(\mu) \equiv \rho_{\rm t}(\mu),
    \label{eq:Bethe-Takahashi}
\end{equation}
where we have denoted by $K(\mu,\nu)$ the scattering kernel of the system. %obtained as the logarithmic derivative of the S-matrix of the corresponding Bethe ansatz solution. 

\begin{figure}[h!]
    \centering
    \includegraphics[width=0.8\columnwidth]{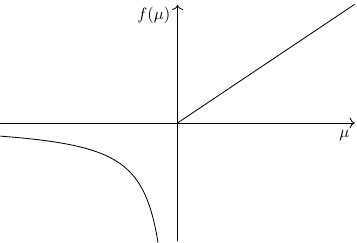}
    \caption{\label{fig:plotF}
    Sketch of a possible reparametrisation $f(\mu)$ that is non-monotonic but injective.}
\end{figure}

The positivity of the driving term guarantees that ${\rho_{\rm t}>0}$, as it is expected for a density. Instead, if the rapidity variable is chosen such that $p(\mu)$ is not monotonic, Eq.~\eqref{eq:Bethe-Takahashi} is not a sensible definition for a density of states. This problem can be treated in several ways, for example by considering separately each branch in which $p(\mu)$ is strictly monotonic, and treating each of such branches as a different quasiparticle species. Here we use a different approach and note that the positivity of momentum can be restored by performing a suitable change of variables $\mu \to f^{-1}(\mu)\equiv \phi$ with $f$ injective (hence invertible) but not monotonic, such as a piecewise function given in Fig.~\ref{fig:plotF}. Indeed, letting $\tilde p(\phi) = p(f(\phi))$, this implies
\begin{equation}
    \frac{d\tilde p}{d\phi} = {p'(f(\phi))}{f'(\phi)}, 
\end{equation}
from which we see that $\tilde p'(\phi)$ can be made strictly positive by choosing a coordinate change, such that $\text{sgn}( p'(f(\phi))) = \text{sgn}(f'(\phi))$. Note that under any rapidity reparametrisation the scattering kernel has to transform in such a way that Eq.~\eqref{eq:Bethe-Takahashi} make $\rho_{\rm t}$ transform correctly as a density. This implies the transformation 
\begin{equation}
    K(\phi,\psi)= f'(\phi) K(f(\phi),f(\psi)) \text{sgn}(f(\psi))\,. 
    \label{eq:kerneltransformationlaw}
\end{equation}

In terms of the new rapidity variable $\phi$, the TBA evaluation can then proceed following standard treatments. In particular, the partition function can be expressed as
\begin{equation}
\begin{aligned}
   & \hspace{-.125cm}\log \mathcal{Z} = \int \frac{d\phi}{2\pi} \tilde p'(\phi) \log(1+\tilde\eta^{-1}(\phi)), \\
   & \hspace{-.125cm}\log\tilde\eta(\phi)=\tilde{\nu} (\phi) + \int \frac{d\phi'}{2\pi} \tilde K(\phi',\phi) \log(1+\tilde\eta^{-1}(\phi')),
\end{aligned}
\label{eq:tbaPhi}
\end{equation}
where $\tilde{\nu}(\phi)$ is the driving term specifying the TBA in the dual channel, expressed in the rapidity variable $\phi$. At this point, it is important to recall that the reparametrisation used is not monotonic. Therefore, if one is to rewrite Eq.~\eqref{eq:tbaPhi} in terms of the original rapidity variables, the sign of the coordinate change would have to be introduced. Since this sign is constructed precisely to be the same as the sign of $p'$, this gives
\begin{equation}
\begin{split}
   & \log \mathcal{Z} = \int \frac{d\mu}{2\pi} p'(\mu) \overline{\mathcal{L}}(\mu), \\
   & \log\overline\eta(\mu)=\tilde \nu(\mu) + s(\mu)  \int \frac{d\mu'}{2\pi}K(\mu',\mu)   \overline{\mathcal{L}}(\mu'),
    \end{split}
    \label{eq:tba_dual_theory}
\end{equation}
where we have used the transformation law for the kernel in Eq.~\eqref{eq:kerneltransformationlaw} together with the transformation of the integration measure, and introduced the shorthand notation
\be
\overline{\mathcal{L}}(\mu) = s(\mu) \log(1+\overline{\eta}^{-1}(\mu)), \,\,\, s(\mu) = \text{sgn}(p'(\mu)). 
\ee
This can be expressed in more convenient form by defining $\eta(\mu) = \overline{\eta}(\mu)^{s(\mu)}$, which satisfies the TBA equations 
\begin{equation}
    \log \eta(\mu) = s(\mu)\tilde  \nu(\mu) +   \int \frac{d\mu'}{2\pi} K(\mu',\mu)  \mathcal{L(\mu')},  
\end{equation}
where the transformation implies that 
\be
\mathcal{L}(\mu) = s(\mu) \log(1+\eta^{-s(\mu)}).
\ee
Note that the occupation functions are naturally defined in terms of $\overline{\eta}(\mu)$ as $\vartheta(\mu) = (1+\overline{\eta}(\mu) )^{-1}$, and therefore when expressed in terms of $\eta(\mu)$ it acquires an extra sign factor
\begin{equation}
    \vartheta(\mu) = (1+\eta(\mu)^{s(\mu)} )^{-1}.
\end{equation}
From the partition function, it is possible to evaluate $d_\alpha = \log \mathcal{Z}_\alpha - \alpha \log \mathcal{Z}$ where $\mathcal{Z}_\alpha$ is the partition function obtained by rescaling the driving term $\nu(\mu)\to \alpha\nu(\mu)$. This can be simplified as
\begin{equation}
    d_\alpha = \int \frac{d\mu}{2\pi} p'(\mu)s(\mu) \log\biggr[(1-\vartheta(\mu)) ^\alpha + \frac{\vartheta(\mu)^\alpha}{x_\alpha(\mu)^{s(\mu)}}\biggl]
\end{equation}
where $x_\alpha = \log\eta_\alpha -\alpha \log\eta$ satisfies
\begin{equation}
    \log x_\alpha(\mu) = \int \frac{d\mu}{2\pi}K(\mu',\mu)s(\mu)\log\biggr[(1-\vartheta(\mu)) ^\alpha + \frac{\vartheta(\mu)^\alpha}{x_\alpha^{s(\mu)}}\biggl].
    \nonumber
\end{equation}

\footnotetext[50]{See the Supplemental Material for: (i) More details on the treatment of inhomogeneous quenches; (ii) Further numerical results.}

\bibliography{bibliography}

\onecolumngrid
%\break
\begin{center}
    {\large \bf Supplemental Material for: \\
           Space-time duality approach to (inhomogeneous) integrable quenches
}

\vspace{0.5cm}

\end{center}
\section{Review of the TBA solution of the XXZ model}
Here we present a minimal self-contained introduction to the thermodynamic properties of the XXZ chain, obtained through the Thermodynamic Bethe Ansatz (TBA) approach, of which a complete review can be found in \cite{takahashi1999book}. Since the theory is gapless for $|\Delta| \leq 1$ and gapped for $\Delta > 1$, leading to considerably different mathematical structures, we treat the two parameter regimes separately in the following subsections. For a more detailed analysis of the gapped case we refer to references \cite{piroli2016, Bertini_2018_lowtransport}, while for the gapless case some more details can be found in the supplementary material of \cite{bertini2023nonequilibrium}.
\subsection{$\Delta > 1$ regime}
The gapped regime is conveniently studied by making use of the parametrisation $\Delta = \cosh\eta$. The general eigenvalue problem can be solved through the Bethe Ansatz method for any system size. Here, however, we focus on the simplifications arising in the thermodynamic limit. 

We consider the system placed on a ring of length $L$ (different boundary conditions are possible, but the standard TBA analyses are performed with PBC), such that
\begin{equation}
    L,N \to \infty, \hspace{0.5cm} N/L  \hspace{0.2cm}\text{ fixed},
\end{equation}
where $N$ the total number of quasiparticles, which in the XXZ model are given by magnons (spin flips above the fully polarised reference state) and their bound states, which are labeled by a parameter $j=1,\dots \infty$. In this limit, the macrostate describing the system is fully described in terms of two quantities, namely the total density of states for each bound state $\rho_{{\rm t},j}$ and the filling function $\theta_j(\mu)$, which specifies the particle density as $\rho_j(\mu) = \theta_j(\mu)\rho_{{\rm t},j}$. While in free theories $\rho_{\rm t}(k) = {1}/{2\pi}$ is completely independent of the filling, in the presence of interactions the two functions are connected through the Bethe-Takahashi equations
\begin{equation}
    \rho_{{\rm t},j}(\mu) = a_j(\mu) - \sum_k \left(K_{jk}*\rho_k\right)(\mu),  
    \label{eq:BTDelta>1}
\end{equation}
where $a*b$ represents the standard convolution and we have introduced the driving term $a_j(\mu) = {p_j'(\mu)}/{2\pi}$ and the scattering kernel $K_{jk}$: these encode the model dependence and in the present context are given by
\begin{equation}
    a_j(\mu) = \frac{1}{\pi}\frac{\sinh(j\eta)}{\cosh(j\eta)-\cosh(2\mu)},\hspace{0.5cm} K_{jk}(\mu) = (1 - \delta_{jk})a_{|j-k|}(\mu) + 2a_{|j-k|+2}(\mu) + \dots + 2a_{j+k-2}(\mu) + a_{j+k}(\mu).
\end{equation}
Equation \eqref{eq:BTDelta>1} alone is not enough to characterize both the total and particle densities within the system. In particular, the filling functions are strongly dependent on the specifics of the protocol: for example, if the system is in a GGE $\sum_n \beta_n Q_n$ these are determined by the generalized TBA equations
\begin{equation}
    \log \frac{1-\theta_j(\mu)}{\theta_j(\mu)} = \sum_n\beta_n q_{n,j}(\mu)\, - \sum_k( K_{jk}* \log(1-\theta_j))(\mu).
    \label{eq:GTBA}
\end{equation}
where $q_{n,j}(\mu)$ are one-particle eigenvalues of the charge operators $Q_n$ on each bound state.
On the other hand, after homogeneous quenches from integrable initial states the the main structure of Eq.~\eqref{eq:GTBA} is preserved, with the only modification being the replacement of $\sum_n \beta_n q_{n,j}(\mu)$ with a different driving term which can be extracted from the overlaps between the initial states and the Hamiltonian's eigenstates~\cite{Caux_2016}. For the standard integrable initial states considered in the literature, namely the tilted N\'eel state 
\begin{equation}
\ket{N,\theta} = e^{i\frac{\theta}{2}\sum_j \sigma_j^x} 
  \ket{\uparrow\downarrow}^{\otimes \frac{L}{2}},
\end{equation}
and tilted ferromagnetic state 
\begin{equation}
\ket{F,\theta} =  e^{i\frac{\theta}{2}\sum_j \sigma_j^x}
  \ket{\uparrow}^{\otimes L},
\end{equation}
and explicit solution of the TBA equations was presented in Ref.~\cite{piroli2016}. Labelling $\theta_j(\mu) = {1}/{(1+\eta_j(\mu))}$, the equations determining $\eta_{n\geq 2}(\mu)$ are the same for the different initial states,
\begin{equation}
    \eta_{n\geq 2} (\mu) = \frac{\eta_{n-1}(\mu + i \eta/2)\eta_{n-1}(\mu - i \eta/2)}{\eta_{n-2}(\mu)+1}-1
\end{equation}
where $\eta_0=0$ and the value of $\eta_1$ is what distinguishes the two states: in particular
\begin{equation}
    \eta_1(\mu) = -1 + \frac{T_1\left(\mu + i \frac{\eta}{2}\right) T_1\left(\mu - i \frac{\eta}{2}\right)}{\phi\left(\mu + i \frac{\eta}{2}\right)\overline\phi\left(\mu - i \frac{\eta}{2}\right)}
\end{equation}
where in the case of the tilted ferromagnet the functions take the form 
\begin{equation}
    \begin{cases}
        T_1(\mu) = \cos(\mu) \left[ 4 \cosh(\eta) - 2 \cos(2\theta) \sin^2\mu + 3 \cos(2\mu) + 1 \right]\\
        \phi(\mu) = 2 \sin^2\theta \sin\mu \cos\left(\mu + i \frac{\eta}{2}\right) \sin\left(\mu - i \frac{\eta}{2}\right) \\
        \overline{\phi}(\mu) = 2 \sin^2\theta \sin\mu \cos\left(\mu - i \frac{\eta}{2}\right) \sin\left(\mu + i \frac{\eta}{2}\right) ,
    \end{cases}
\end{equation}
while for the tilted Nèel state 
\begin{equation}
\begin{cases}
T_1(\mu) = -\frac{1}{8} \cot(\mu) \left[ 8 \cosh(\eta) \sin^2(\theta) \sin^2(\mu) - 4 \cosh(2\eta) \right.\\ \left. \hspace{1.4cm} + (\cos(2\theta) + 3)(2 \cos(2\mu) - 1) + 2 \sin^2(\theta) \cos(4\mu) \right] \\
\phi(\mu) = \frac{1}{8} \sin(2\mu + i\eta) \left[ 2 \sin^2(\theta) \cos(2\mu - i\eta) + \cos(2\theta) + 3 \right]  \\
\bar{\phi}(\mu) = \frac{1}{8} \sin(2\mu - i\eta) \left[ 2 \sin^2(\theta) \cos(2\mu + i\eta) + \cos(2\theta) + 3 \right] .
\end{cases}
\end{equation}

Beyond the description of the macrostate, the TBA framework also allows to characterize the fundamental excitations on top of it, which are responsible for the large space-time scale dynamics. In particular, the presence of interactions is mainly encoded in a renormalization of the energy and momentum of each quasiparticle, through the ``dressing'' equation
\begin{equation}
    f_j^{\rm dr}(\mu)= f_j(\mu) - \sum_k K_{jk}*(\theta_k f_k^{dr})(\mu).
\end{equation}
In this light of this, the TBA solution can be interpreted as the thermodynamics arising from a set of free quasiparticles characterized by momentum $p_j^{\rm dr}$ and energy $\varepsilon_j^{\rm dr}$, propagating at a velocity 
\begin{equation}
    v^{\rm dr}_j(\mu) = \frac{\partial\varepsilon^{\rm dr}_j(\mu)}{\partial p_j^{\rm dr}(\mu)} = \frac{(\varepsilon')^{\rm dr}(\mu)}{2\pi \rho_{{\rm t},j}(\mu)} \implies 2\pi\, v^{\rm dr}_j(\mu)\rho_{{\rm t},j}(\mu)=(\varepsilon')^{\rm dr}(\mu)\,,
\end{equation}
where we have used that $(p_j')^{dr}=2\pi \rho^t_j$, as implicit in the Bethe-Takahashi equations.
% Note, however, that the failure of the quasiparticle picture implies that it is not strictly true that the effect of the interactions on the state is fully encoded in a simple renormalization of the energy and momentum of excitations, and this statement is only true in the evaluation of the densities; this is the case for example in the context of GHD, which will be reviewed in the following section. In the evaluation of more complicated quantities, including the Rènyi entropies, the action of the dressing appears in a more subtle and nontrivial way, as shown in the main result \eqref{eq:slopeSDA}.

\subsection{$\Delta \leq 1$ at roots of unity}
In the gapless regime, we focus on a specific set of parameter values, for which the TBA treatment admits a considerable simplification,
\begin{equation}
    \Delta = \cos\eta, \hspace{0.5cm} \eta = \frac{\pi}{p+1}, \hspace{0.1cm} p\in\mathbb{N}.
\end{equation}
It is possible to show that this parameter choice leads to a truncation in the number of bound states; in particular, the total number of quasiparticle species is $p+1$. Defining the string parity and string length as
\begin{equation}
    \nu_j = \begin{cases}
        1 \hspace{0.8cm}  \text{for } j \leq p \\
        -1  \hspace{0.5cm}  \text{for } j =p+1
    \end{cases} \hspace{1cm}q_j = \begin{cases}
        j \hspace{0.8cm}  \text{for } j \leq p \\
        1  \hspace{0.8cm}  \text{for } j =p+1 \, ,
    \end{cases}
\end{equation} 
and introducing the auxiliary function
\begin{equation}
    \tilde a^\nu_n(\mu) = \frac{\nu \sin\frac{n\pi}{p+1}}{\pi\left(\cosh(2\mu) - \nu\cos \frac{\pi n}{p+1}\right)},
\end{equation}
it is possible to define the driving term for the Bethe-Takahashi equations as 
\begin{equation}
    a_j(\mu) = \tilde a^{\nu_j}_{q_j}(\mu),
\end{equation}
which gives 
\begin{equation}
    \rho_{{\rm t},j}(\mu) = \nu_j a_j(\mu) - \nu_j\sum_{k=1}^{p+1} (K_{jk} *\rho_j)(\mu).
\end{equation}
In this case the scattering kernel $K_{jk}$ has the slightly more involved expression
\begin{equation}
    K_{jk} = (1 - \delta_{q_j, q_k}) \tilde{a}_{|q_j - q_k|}^{\nu_j \nu_k}(\mu) + 2 \left( \tilde{a}_{|q_j - q_k| + 2}^{\nu_j \nu_k}(\mu) + \tilde{a}_{|q_j - q_k| + 4}^{\nu_j \nu_k}(\mu) + \dots + \tilde{a}_{q_j + q_k - 2}^{\nu_j \nu_k}(\mu) \right) + \tilde{a}_{q_j + q_k}^{\nu_j \nu_k}(\mu).
\end{equation}
Again, the filling functions contain the dependence on the initial state determining the quench, and can be determined case by case. Introducing the auxiliary functions
\begin{equation}
    s(\mu) = \frac{p + 1}{2\pi \cosh \left( (p + 1)\mu \right)},\hspace{0.5cm} d(\mu) = \log \left[ \coth \left( \frac{(p + 1)\mu}{2} \right) \right]^2,
\end{equation}
in the case of a quench from the N\'eel state the eta functions $\eta_j(\mu)$ are determined as
\begin{equation}
\begin{cases}
\log \eta_1 &= (1 + \delta_{p,2}) s \star \log(1 + \eta_2) - d(\lambda), \\
\log \eta_n &= s \star \left[ \log(1 + \eta_{n-1}) + (1 + \delta_{p,n+1}) \log(1 + \eta_{n+1}) \right] + (-1)^n d(\lambda), \\
\log \eta_p &= s \star \log(1 + \eta_{p-1}) + d(\lambda) \delta_{p (\text{mod}2),0}\, ,  \\
\log \eta_{p+1} &= -\log \eta_p
\end{cases}
\end{equation}
where $1<n<p$.  Another interesting initial state is the Dimer (or Majumdar-Ghosh) state 
\begin{equation}
    \ket{D} = \frac{1}{\sqrt{2}}\left[  (|\uparrow\downarrow\rangle - |\downarrow\uparrow\rangle) \right]^{\otimes L/2}
\end{equation}
in which the eta functions can be characterised by a similar set of equations
\begin{equation}
\begin{cases}
\log \eta_1 &= (1 + \delta_{p,2}) s \star \log(1 + \eta_2) - d(\lambda), \\
\log \eta_n &= s \star \left[ \log(1 + \eta_{n-1}) + (1 + \delta_{p,n+1}) \log(1 + \eta_{n+1}) \right] - d(\lambda), \\
\log \eta_p &= s \star \log(1 + \eta_{p-1}) - d(\lambda), \\
\log \eta_{p+1} &= -\log \eta_p
\end{cases}.
\end{equation}

\section{Bipartite quenches}
In this section we specify the results of the main text to the typical scenario of the bipartitioning protocol, in which the GHD equation admit an exact solution. For simplicity, we restrict to the case of a theory with a single quasiparticle species, the generalisation to multiple species is straightforward. In inhomogeneous setups, we build a hydrodynamic framework by subdividing the system in mesoscopic fluid cells, and consider Eq.~\eqref{eq:slopeSDA} to be valid at the level of each cell, with space-time dependence entering through the occupation functions $\theta(\mu,x,t)$ defining the state.
In the bipartite scenario, the scale invariance of the system implies that the GHD solution for the occupation functions is constant along fixed rays $\zeta=x/t$, hence giving $\theta(\mu,x,t) =  \theta(\mu,\zeta)$. 

We first recall some basic notions of Generalized Hydrodynamics which are needed to proceed (see, e.g., Ref.~\cite{Alba_2021}). Firstly, the occupation functions $ \theta(\mu,\zeta')$ satisfy the convective equation  
\begin{equation}
     \partial_x \theta(\mu,x,t) + v^{\rm dr}(\mu,x,t) \partial_t \theta(\mu,x,t) = 0   \implies  v^{\rm dr}_{\zeta}(\mu)   \partial_\zeta   \theta(\mu,\zeta) = \zeta  \partial_\zeta   \theta(\mu,\zeta).
     \label{eq:GHD_occupation}
\end{equation}
 Note that an equation of the form like that in Eq.~\eqref{eq:GHD_occupation} holds for any function of $ \theta(\mu,\zeta)$, such that $v^{\rm dr}_{\zeta}   \partial_\zeta  F( \theta(\mu,\zeta)) = \zeta  \partial_\zeta  F( \theta(\mu,\zeta))$. Moreover, the particle and root densities satisfy continuity equations, reflecting the fact that both are conserved in integrable models, which in terms of the rays are expressed as
   \begin{equation}
       \partial_\zeta (v^{\rm dr}_{\zeta}   \rho_{\rm t,\zeta} ) = \zeta  \partial_\zeta  \rho_{{\rm t},\zeta}, \hspace{0.5 cm} \partial_\zeta (v^{\rm dr}_{\zeta}   \rho_\zeta ) = \zeta  \partial_\zeta  \rho_\zeta,
       \label{eq:GHD_continuity}
   \end{equation}
where the rapidity dependence is left implicit. For a bipartite quench, the GHD equations allow for an exact implicit solution  for the filling functions in terms of the occupations $\theta_L(\mu)$ and $\theta_R(\mu)$ of the two reservoirs
\begin{equation}
    \theta(\mu,\zeta) = \theta_{\text{L}}(\mu) \Theta\big( v_\zeta^{\text{dr}}( \mu) - \zeta \big) + \theta_{\text{R}}(\mu) \Theta\big( \zeta - v^{\text{dr}}_\zeta(\mu) \big).
    \label{bipartite_occupation}
\end{equation}
Note that this is an implicit solution, as the dressed velocity depends itself on the value of the filling function. This expression allows, however, for a very efficient numerical evaluation.

We are now in a position to continue with our discussion of SDA results for bipartite quenches. The key idea is that each ray $\zeta$ contributes a factor which is precisely the density of entanglement at the given ray in the swapped theory, given by \eqref{eq:slopeSDA} specialized to the inhomogeneous values for $\rho_{\rm t,\zeta}$ and $\theta_\zeta$ obtained through the GHD solution. This allows us to write 
\begin{equation}
\begin{split}
     s_\alpha(\zeta)=  \log\mathcal{Z}_\alpha^{(\zeta)} - \alpha \log \mathcal{Z}^{(\zeta)}= \hspace{-0.1cm}\int \frac{d\mu}{2\pi}  v^{\rm dr}_{\zeta'}(\mu)\rho_{t,{\zeta'}} \Biggl\{ \theta(\mu,\zeta') \log y_{{\zeta'},\alpha} +   s(\mu)\log\left((1- \theta(\mu,\zeta'))^\alpha + \frac{\theta^\alpha(\mu,\zeta')}{y_{{\zeta'},\alpha}^{s(\mu)}}\right)  \Biggl\},
\end{split}
\end{equation}
where we have expressed the local entropy slopes in terms of partition functions in the language of \cite{alba_quench_action_2017,bertini2022growth}, see also the End Matter.
The leading order solution for the full entropy is obtained by integrating over such contributions in the time interval shown in Fig.~\ref{fig:zetaneq0} of the main text. This gives
\begin{equation}
\begin{split}
    &S^{(\alpha)}_{[x,\infty]}(t) = \frac{1}{1-\alpha}\int_0^t dt'  \left(\log \mathcal{Z}^{(\zeta)}_\alpha - \alpha \log \mathcal{Z}^{(\zeta)}\right) = \frac{x}{1-\alpha}\int_{\zeta}^\infty \frac{d\zeta' }{(\zeta') ^2}  \left(\log \mathcal{Z}^{(\zeta' )}_\alpha - \alpha \log \mathcal{Z}^{(\zeta' )}\right). 
\end{split}
\end{equation}
Considering the first term we have 
\begin{eqnarray}
 - x \int_{\zeta}^\infty \frac{d\zeta' }{(\zeta') ^2}   \log \mathcal{Z}^{(\zeta' )} &=& x \int_{\frac{x}{t}}^\infty \frac{d\zeta' }{(\zeta') ^2}   \int \frac{d\mu}{2\pi} \varepsilon'(\mu) s(\mu) \log(1-\theta(\mu,\zeta' ))\\ &=& x \int d\mu s(\mu) \int_{x/t}^{\infty} \frac{d\zeta' }{(\zeta') ^2} v^{\rm dr}_{\zeta'}  \rho_{t,\zeta' } \left\{\log(1-\theta(\mu,\zeta' )) + \theta(\mu,\zeta' ) (\tilde \nu_{\zeta}-\log \overline \eta_{\zeta'}) \right\}, 
\label{integraltosolve}
\end{eqnarray}
where in the second line we have used the relation between dressed velocity, total density and dressing of the derivative of the energy discussed in the previous section, i.e.,  
\begin{equation}
    \frac{\varepsilon'(\mu)}{2\pi} = v^{\rm dr}_{\zeta'}   \rho_{\rm t,\zeta' } + \int d\mu K(\mu,\mu')  \theta(\mu,\zeta')v^{\rm dr}_{\zeta'}  \rho_{\rm t,\zeta' },
    \label{eq:dressedmomentum=vrho}
\end{equation}
together with Eq.~\eqref{eq:tba_dual_theory} for the TBA in the dual theory (defined by the driving term $\tilde \nu_\zeta$ which is fixed by the requirement that the occupation functions are the same in the two theories). For notational convenience we will refer to the term in the curly bracket as $F( \theta(\mu,\zeta'))$. Using the GHD identities discussed above, Eq.~\eqref{integraltosolve} can be solved by partial integration, giving 
   \begin{equation}
      x  \int_{\zeta}^{\infty} \frac{d\zeta' }{(\zeta') ^2} v^{dr}_{\zeta'}   \rho_{t,\zeta' } F( \theta(\mu,\zeta')) = -x\left[\frac{1}{\zeta' }v^{\rm dr}_{\zeta'}   \rho_{t,\zeta' } F( \theta(\mu,\zeta'))\right]_{\zeta}^\infty + x \int_{\zeta}^{\infty} \frac{d\zeta' }{\zeta' }\partial_{\zeta'} \left( v^{\rm dr}_{\zeta'}   \rho_{t,\zeta' } F( \theta(\mu,\zeta'))\right).
   \end{equation}
The last term is simplified by making use of Eqs.~\eqref{eq:GHD_occupation} and \eqref{eq:GHD_continuity}, i.e. 
\begin{equation}
   \partial_{\zeta'} \left( v^{\rm dr}_{\zeta'}   \rho_{t,\zeta' } F( \theta(\mu,\zeta'))\right) =\zeta'  \partial_{\zeta'} (\rho_{t,\zeta' } F( \theta(\mu,\zeta'))),
\end{equation}
allowing us to simplify the integral to 
\begin{equation}
     x  \int_{x/t}^{\infty} \frac{d\zeta' }{(\zeta') ^2} v^{\rm dr}_{\zeta'}   \rho_{t,\zeta' } F( \theta(\mu,\zeta'))  = t(v^{\rm dr}_{\zeta} - \zeta) \rho_{\rm t,\zeta} F( \theta(\mu,\zeta)).   
\end{equation}
Finally, putting all together we obtain 
\begin{equation}
-x \int_{\zeta}^\infty \frac{d\zeta' }{(\zeta') ^2}   \log \mathcal{Z}^{(\zeta' )} = t \int d\mu s(\mu) (v^{\rm dr}_{\zeta} - \zeta) \rho_{\rm t,\zeta} F( \theta(\mu,\zeta)) =  t \int \frac{d\mu}{2\pi} s(\mu) (\varepsilon'(\mu) - \zeta p'(\mu))  \log(1- \theta(\mu,\zeta)), 
\end{equation}
where we have used Eq.~\eqref{eq:dressedmomentum=vrho} and $(p')^{\rm dr}_{\zeta}=2\pi \rho_{\rm t,\zeta}$. Note that an identical discussion can be performed for the term $\log \mathcal{Z}_\alpha$: in this case, it is possible to view it as the partition function of a model with driving term $\alpha \tilde\nu_{\zeta}$, and, therefore, to define $\alpha$-dependent effective velocity and occupation function fulfilling Eqs.~\eqref{eq:GHD_occupation} and \eqref{eq:GHD_continuity}. This allows us to conclude
\begin{equation}
   (1-\alpha)  S^{(\alpha)}_{[x,\infty]}(t) =  t \int \frac{d\mu}{2\pi} s(\mu) (\varepsilon'(\mu) - \zeta  p'(\mu))  \left(\alpha \log(1- \theta(\mu,\zeta)) - \log(1-\theta^{(\alpha)}_\zeta(\mu))\right),  
\end{equation}
where the integrand can be rewritten as 
\begin{equation}
    \alpha \log(1- \theta(\mu,\zeta)) - \log(1-\theta^{(\alpha)}_\zeta(\mu)) = \log\left((1- \theta(\mu,\zeta))^\alpha + \frac{\theta^\alpha_\zeta(\mu)}{y_{\zeta,\alpha}^{s(\mu)}}\right),
\end{equation}
with the auxiliary function $y_{\zeta,\alpha}$ satisfying a non-trivial ray dependent dressing
\begin{equation}
    \log y_{\zeta,\alpha} = \int d\mu' K(\mu,\mu') s(\mu')  \log\left((1- \theta(\mu,\zeta))^\alpha + \frac{\theta^\alpha_\zeta(\mu)}{y_{\zeta,\alpha}^{s(\mu)}}\right),
\end{equation}
reproducing the form discussed in the main text. Exploiting the latter equation, $S^{(\alpha)}_{[x,\infty]}(t)$ can be further rewritten in a way that is closer in spirit to the formulation of Ref.~\cite{alba2019entanglement}  
\begin{equation}
    S^{(\alpha)}_{[x,\infty]}(t) =  t  \int d\mu\, s(\mu) (v^{\rm dr}_\zeta(\mu) - \zeta)\rho_{\rm t,\zeta }\left\{\log\left((1- \theta(\mu,\zeta))^\alpha + \frac{\theta^\alpha_\zeta(\mu)}{y_{\zeta,\alpha}^{s(\mu)}}\right) + s(\mu)  \theta(\mu,\zeta) \log y_{\zeta,\alpha} \right\}. 
\end{equation}
Note that the solution for a generic ray $\zeta$ is essentially the same as the one for the ray $\zeta=0$ after moving to the comoving frame. Moreover, since in the $\alpha \to 1$ limit $\log y_{\zeta,\alpha}^{s(\mu)}$ vanishes just as in the homogeneous case, we see that in the Von Neumann limit the result reduces to 
\begin{equation}
     S_{[x,\infty]}(t) =  -t  \int d\mu s(\mu) (v^{dr}_\zeta(\mu) - \zeta)\rho_{\rm t,\zeta}\left[(1- \theta(\mu,\zeta))\log(1- \theta(\mu,\zeta)) +  \theta(\mu,\zeta) \log \theta(\mu,\zeta)\right].
\end{equation}
which is the quasiparticle-picture formula proposed in Ref.~\cite{alba2019entanglement}. 

Note that this also gives the correct prediction in the homogeneous limit: in that case the term in multiplying the global factor $\zeta$ vanishes (the filling functions are even for integrable states) giving  
\begin{equation}
      S^{(\alpha),\rm hom}_{[x,\infty]}(t) =  t  \int d\mu s(\mu) v^{dr}_\zeta(\mu)\rho_{\rm t,\zeta}\left\{\log\left((1- \theta(\mu,\zeta))^\alpha + \frac{\theta^\alpha_\zeta(\mu)}{y_{\zeta,\alpha}^{s(\mu)}}\right) + s(\mu)  \theta(\mu,\zeta) \log y_{\zeta,\alpha} \right\}
\end{equation}
which corresponds to the result of \cite{bertini2022growth}. As a final comment, we note that for values of $|\zeta|$ which are larger than the maximum possible value of the dressed velocity, the solution reduces to the homogeneous one with occupation function $\theta_L(\mu)$ or $\theta_R(\mu)$ depending on wether $x<0$ or $x>0$ respectively. This is a natural implication of Lieb Robinson bounds, which imply that the nontrivial physics takes place only within the light cone. 

The discussion regarding the FCS of the current can be carried on similar grounds. First one assigns the contribution in Eq.~\eqref{eq:FCSSDA} of the main text to each fluid cell and then integrates over the corresponding time interval. Manipulating the integral by directly repeating the procedure presented here leads to Eq.~\eqref{eq:FCSgeneralzeta} of the main text.

\section{Details on the numerical analysis}

The numerical evaluation is performed implementing the TEBD scheme by means of the iTensor Julia library~\cite{iTensor}. Representing the initial state as a Matrix Product State (MPS) the time evolution is approximated via a Trotter-Suzuki decomposition of the time evolution operator, which can be taken as the forward-backward swift
\begin{equation}
    {\rm{U}}(dt) = \prod_{i=0}^{L} e^{-ih_{i,i+1} dt}  \prod_{i=L}^{0} e^{-ih_{i,i-1} dt}
\end{equation}
where each term $e^{-ih_{i,i+1} dt} $ is considered as a quantum gate acting on two spins. This produces errors of order $dt^3$ per time step, and allows us to obtain good numerical results for $dt = 0.01-0.1$. We consider a global system of size $L=80$ with open boundary conditions, and a subsystem of size $L_A=40$ placed on the left half of the system. 

Truncations are performed to avoid an excessive growth of the MPS. In particular, we use a cutoff of $10^{-12}$ on the Schmidt values at each step and a maximum bond dimension $\chi=1024$: these allow us to reach values of $t\sim 10$ by keeping a good precision in the description of the state, with variations depending on the entanglement slope for each quench.

\begin{figure}[t]
    \begin{subfigure}[c]{0.48\textwidth}
        \centering
        \includegraphics[width=\linewidth]{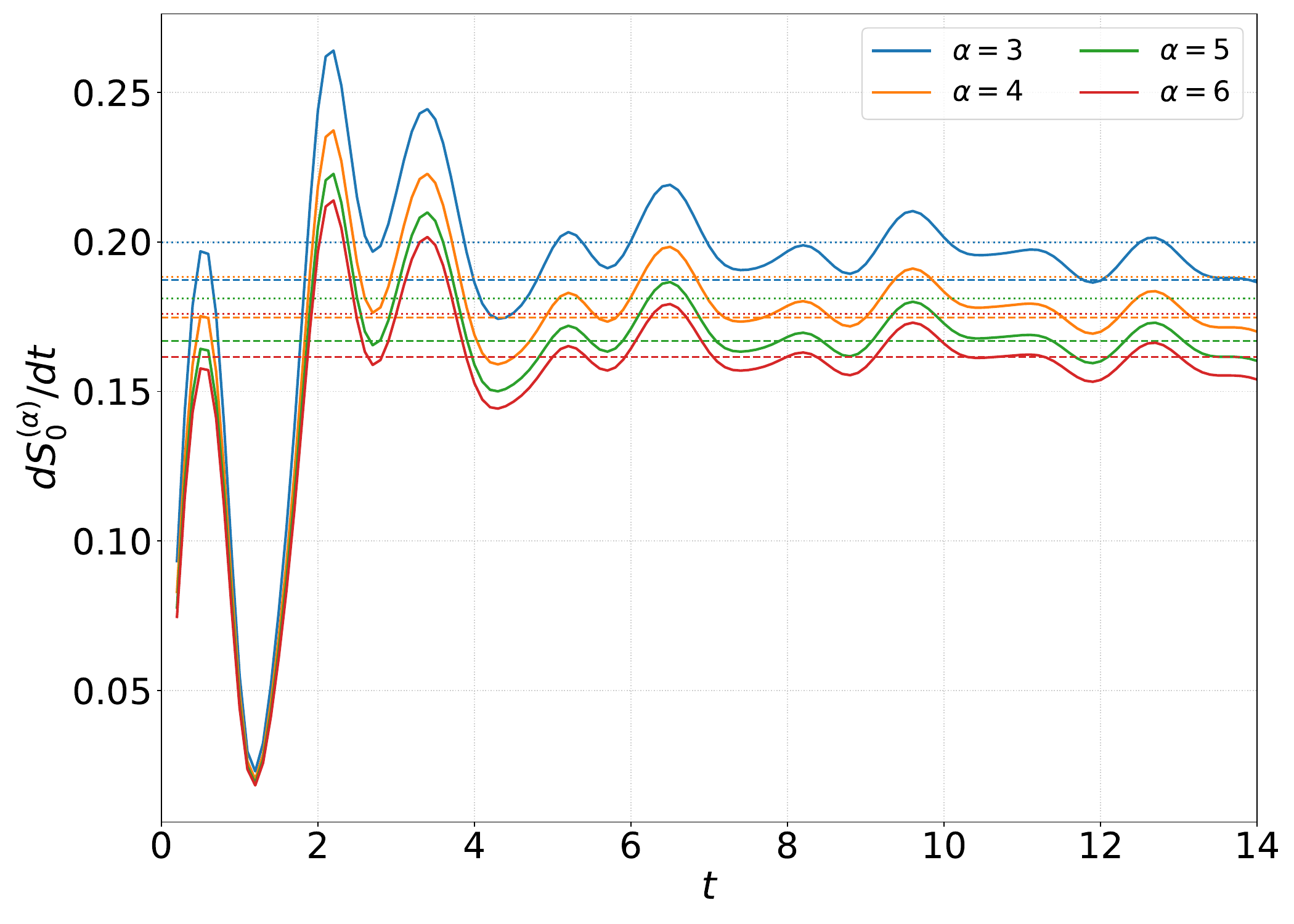}
    \end{subfigure}
    \hfill
    \begin{subfigure}[c]{0.48\textwidth}
        \centering
        \includegraphics[width=\linewidth]{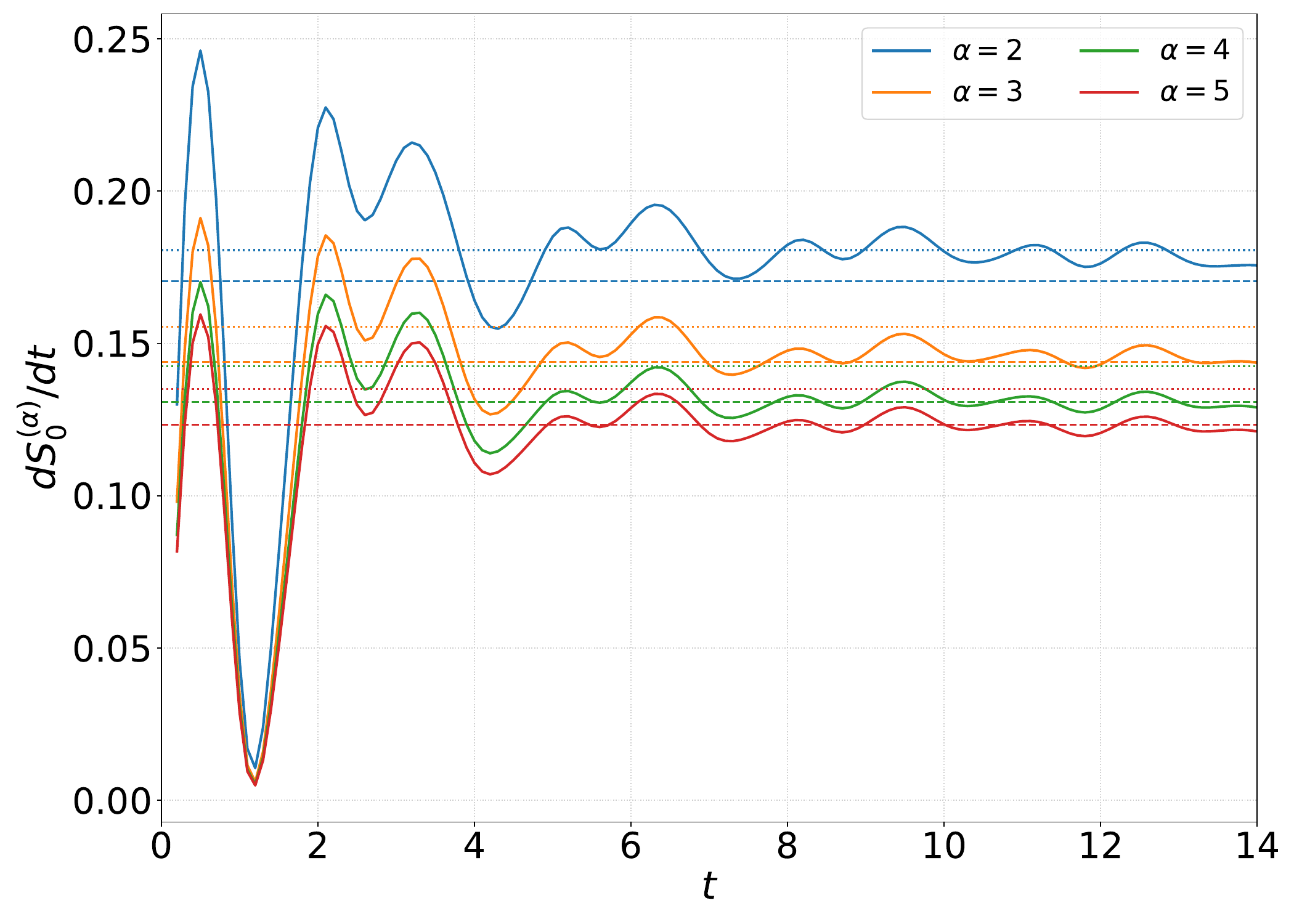}
    \end{subfigure}
    \caption{\label{fig:comparison1}
      Instantaneous slope of Rényi entropies in two quenches in the XXZ chain with $\Delta=4$. On the left the initial state is $\ket{N,0}_L\otimes\ket{N,{\pi}/{6}}_R$, while on the right $\ket{N,0}_L\otimes\ket{N,{\pi}/{8}}_R$. Dashed lines represent the predictions obtained by using $\text{sgn}(\varepsilon'(\theta))$, while dotted lines are the predictions obtained using $\text{sgn}(v^{\rm dr}(\theta))$. In both cases, while for lower $\alpha$ it is impossible to distinguish between the two predictions at the scale accessible via TEBD, the solution at higher $\alpha$ shows clearly that the numerical solution converges to the dashed line.}
\end{figure}

\begin{figure}[h!]
    \begin{subfigure}[c]{0.48\textwidth}
        \centering
        \includegraphics[width=\linewidth]{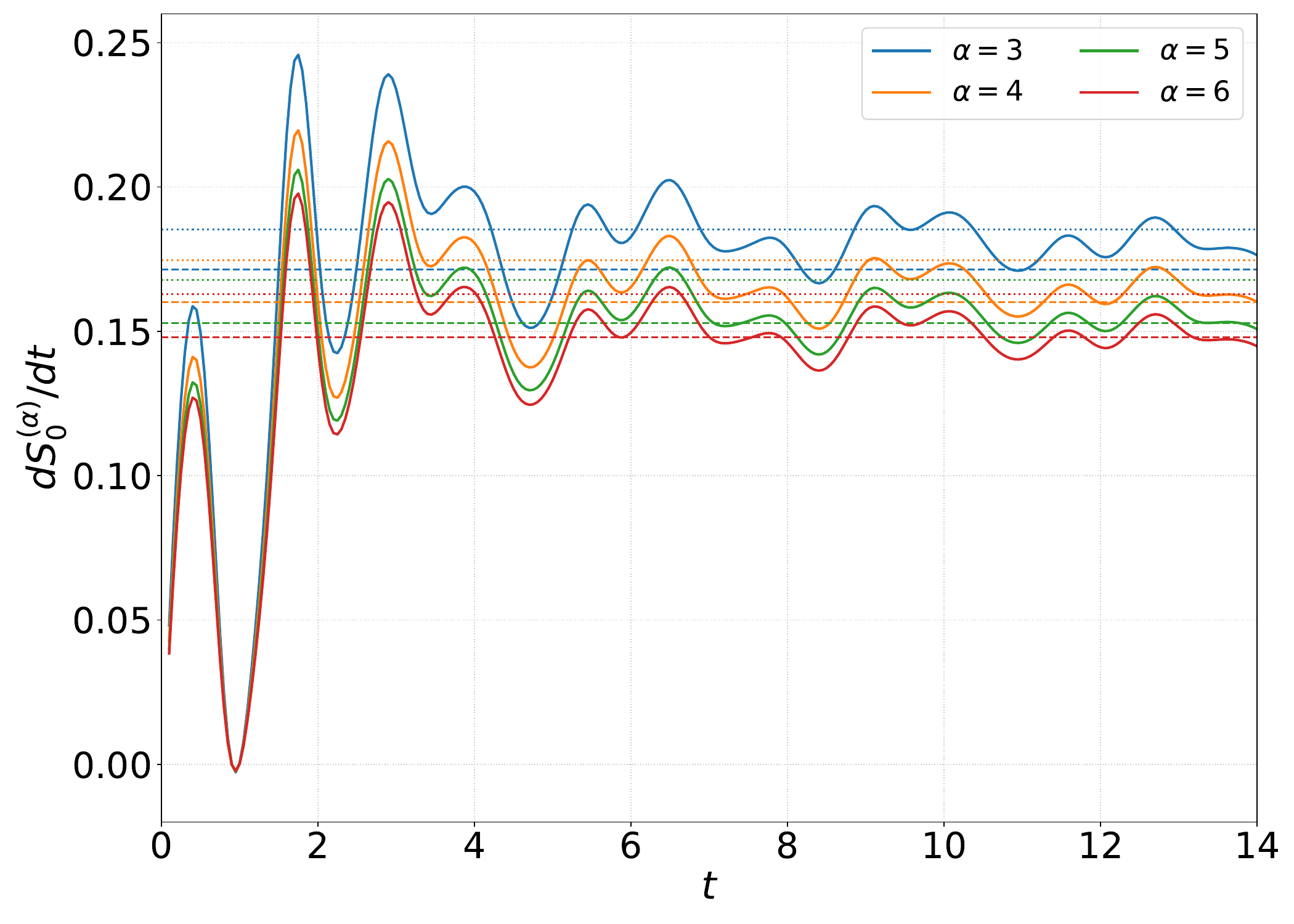}
    \end{subfigure}
    \hfill
    \begin{subfigure}[c]{0.48\textwidth}
        \centering
        \includegraphics[width=\linewidth]{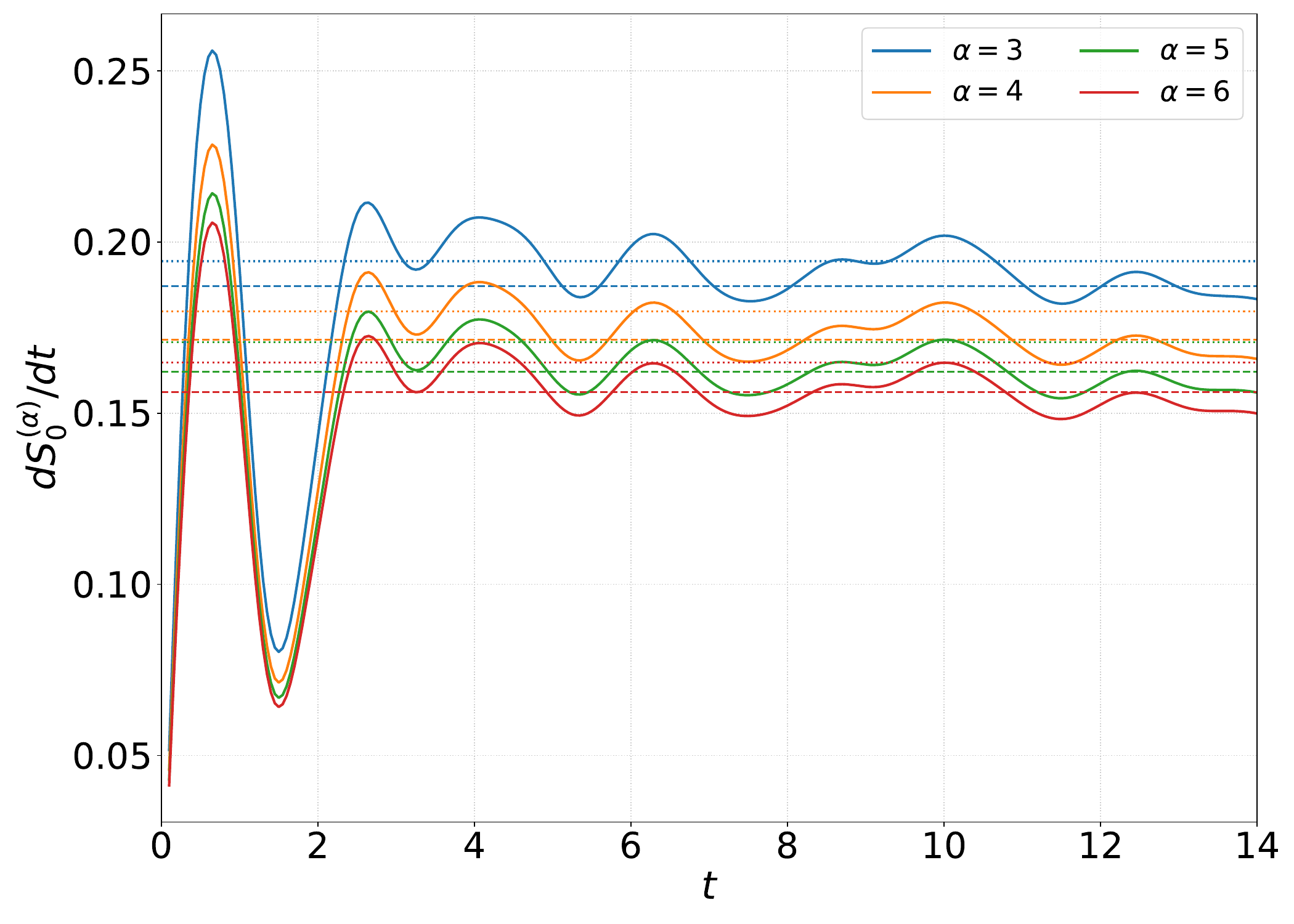}
    \end{subfigure}
    \caption{\label{fig:comparison2}
      Instantaneous slope of Rényi entropies in two quenches in the XXZ chain with $\Delta=5$ (on the left) and $\Delta =3$ (on the right). On the left the initial state is $\ket{N,0}_L\otimes\ket{N,{\pi}/{6}}_R$, while on the right $\ket{N,0}_L\otimes\ket{N,{\pi}/{8}}_R$. Dashed lines represent the predictions obtained by using $\text{sgn}(\varepsilon'(\theta))$, while dotted lines are the predictions obtained using $\text{sgn}(v^{\rm dr}(\theta))$. In both cases, while for lower $\alpha$ it is impossible to distinguish between the two predictions at the scale accessible via TEBD, the solution at higher $\alpha$ shows clearly that the numerical solution is closer to the dashed line.}
\end{figure}

\begin{figure}[ht!]
    % Row 2
    \begin{subfigure}[c]{0.48\textwidth}
        \centering
        \includegraphics[width=\linewidth]{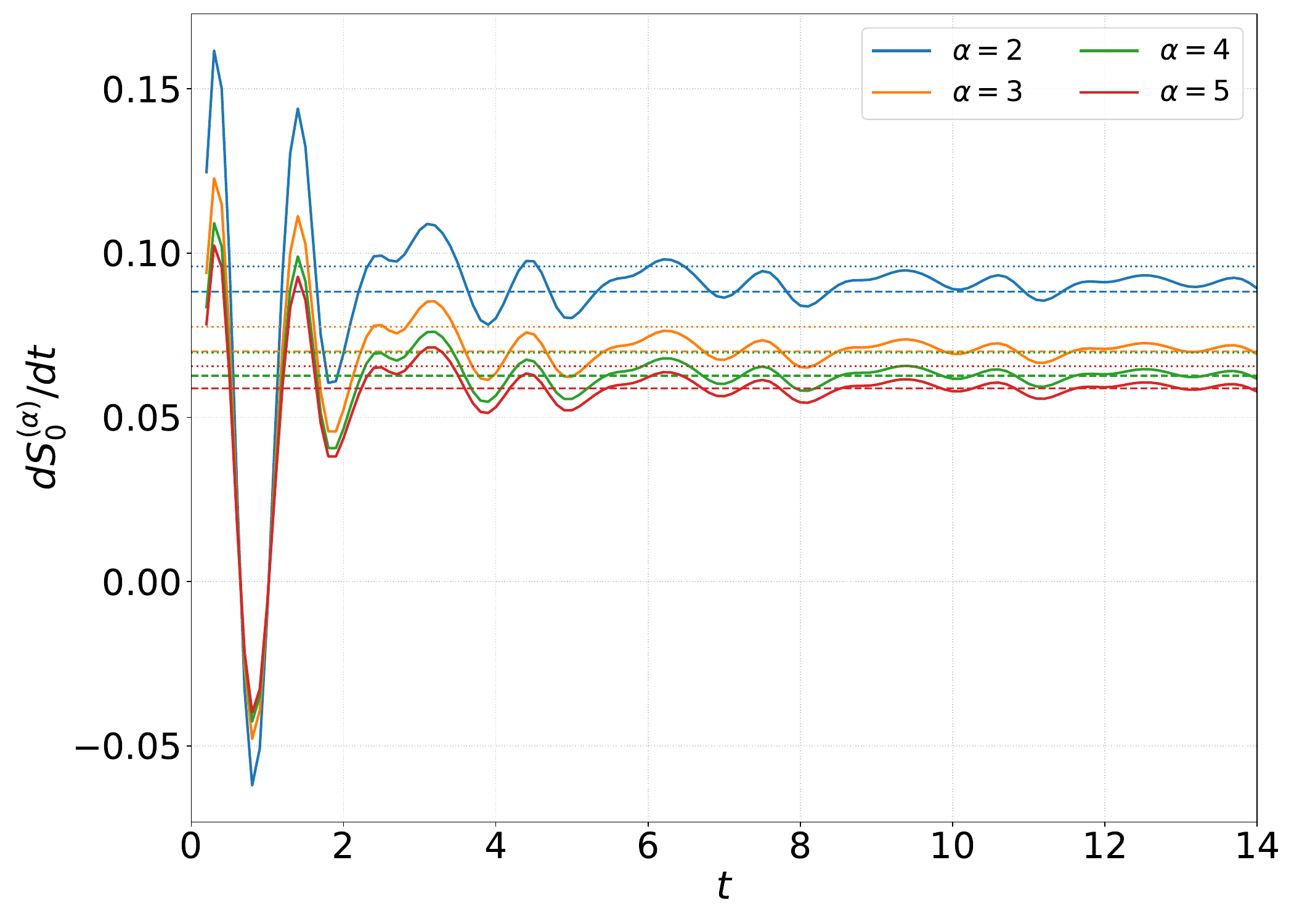}
    \end{subfigure}
    \hfill
    \begin{subfigure}[c]{0.48\textwidth}
        \centering
        \includegraphics[width=\linewidth]{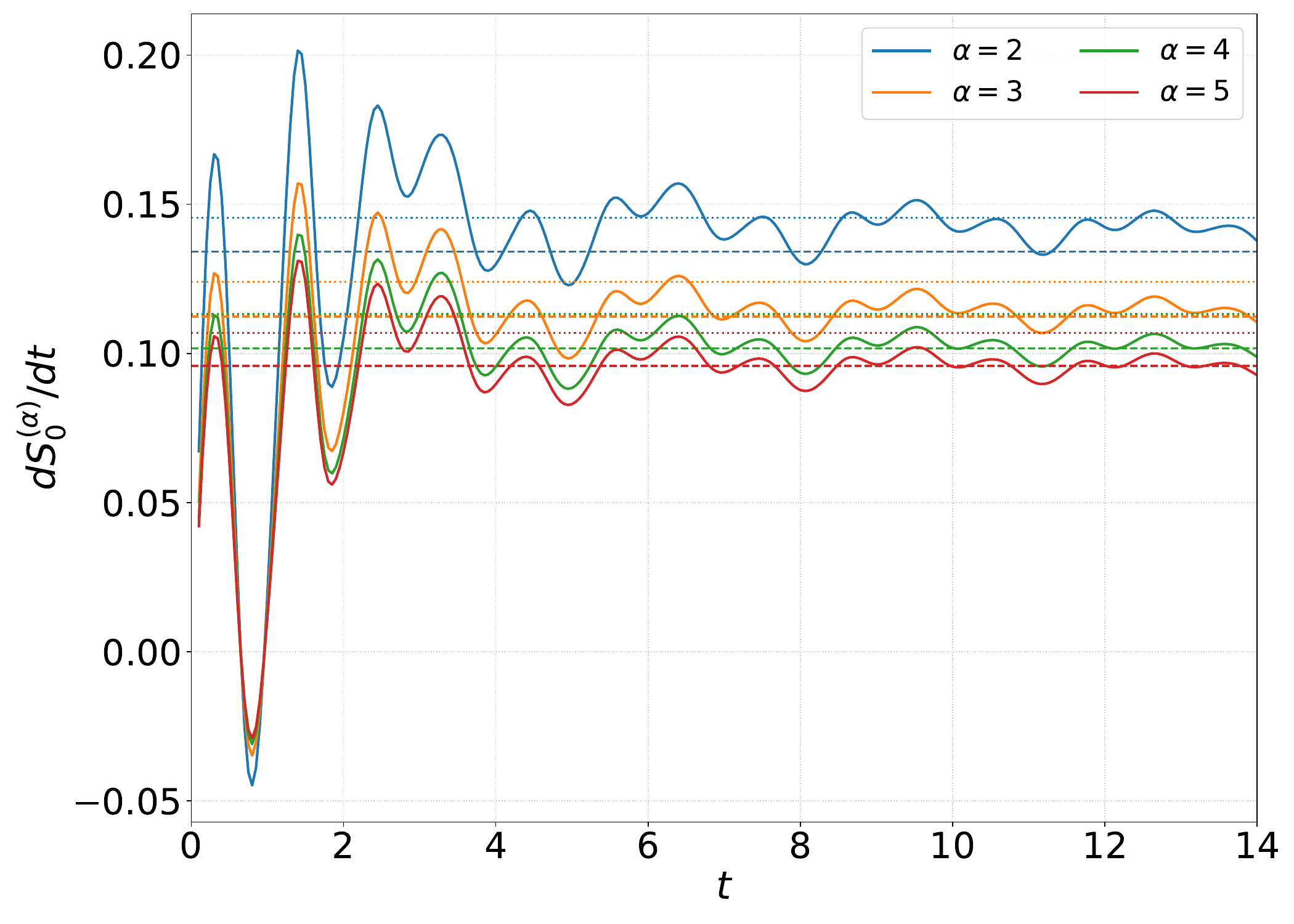}
    \end{subfigure}
    \caption{\label{fig:freeplots2}
      Instantaneous slope of Rényi entropies in two quenches in the XXZ chain with $\Delta =6$. On the left  the initial state is $\ket{N,0}_L\otimes\ket{N,{\pi}/{12}}_R$, while on the right is $\ket{N,0}_L\otimes\ket{N,{\pi}/{8}}_R$. Dashed lines represent the predictions obtained by using $\text{sgn}(\varepsilon'(\theta))$, while dotted lines are the predictions obtained using $\text{sgn}(v^{\rm dr}(\theta))$. The plot on the left corresponds to the same initial state as Fig.~\ref{fig:comparisonpi12} in the main text, for a higher value of $\Delta$. We see that increasing $\Delta$ leads to better agreement between TEBD and the prediction at comparable values of time.}
\end{figure}

The cutoff of smaller Schmidt values has the consequence that the lower the Rényi index, the larger will be the cumulative effect of the truncation in time. This implies that larger values of $\alpha$ allow to minimize such numerical effects, leading to more reliable numerical results. 
Moreover, it is convenient to consider the instantaneous slope ${\rm d}S_0^{(\alpha)}/{\rm d} t$, rather than the global one $S_0^{(\alpha)}(t)/t$. This is useful because the global slope is affected by the large spike present in the initial time regime, which eventually gets canceled on time scales that are larger than those accessible in the TN evaluation. The TEBD result is then compared with the analytical predictions obtained in the main text. Representative examples of our results for the slope of R\'enyi entropies are presented in Figs.~\ref{fig:comparison1}, \ref{fig:comparison2}, and \ref{fig:freeplots2}, as well as Fig.~\ref{fig:comparisonpi12} of the main text. The figures compare the TEBD results with our prediction in Eq.~\eqref{eq:Sgeneralzeta} of the main text evaluated at $\zeta=0$. For reference we also report the prediction obtained by replacing ${\rm sgn}(\varepsilon'(\mu))$ with ${\rm sgn}(v^{\rm dr}(\mu))$ in in Eq.~\eqref{eq:Sgeneralzeta}, i.e.\ deliberately missing the effect of curved quasiparticle trajectories. The results at higher $\alpha$ clearly show convergence to the prediction of Eq.~\eqref{eq:Sgeneralzeta} rather than to the modified one.  

The evaluation of the FCS can be performed in a similar manner. In this case the initial state is written as 
\begin{equation}
    e^{-i\beta Q_A(0)} \ket{\psi_0}.
\end{equation}  
Since the bond dimension of the operator is just $\chi_{MPO}=1$, this does not appreciably increase the complexity of the evolution through TEBD. 

\begin{figure}[h!]
    % Row 2
    \begin{subfigure}[c]{0.48\textwidth}
        \centering
        \includegraphics[width=\linewidth]{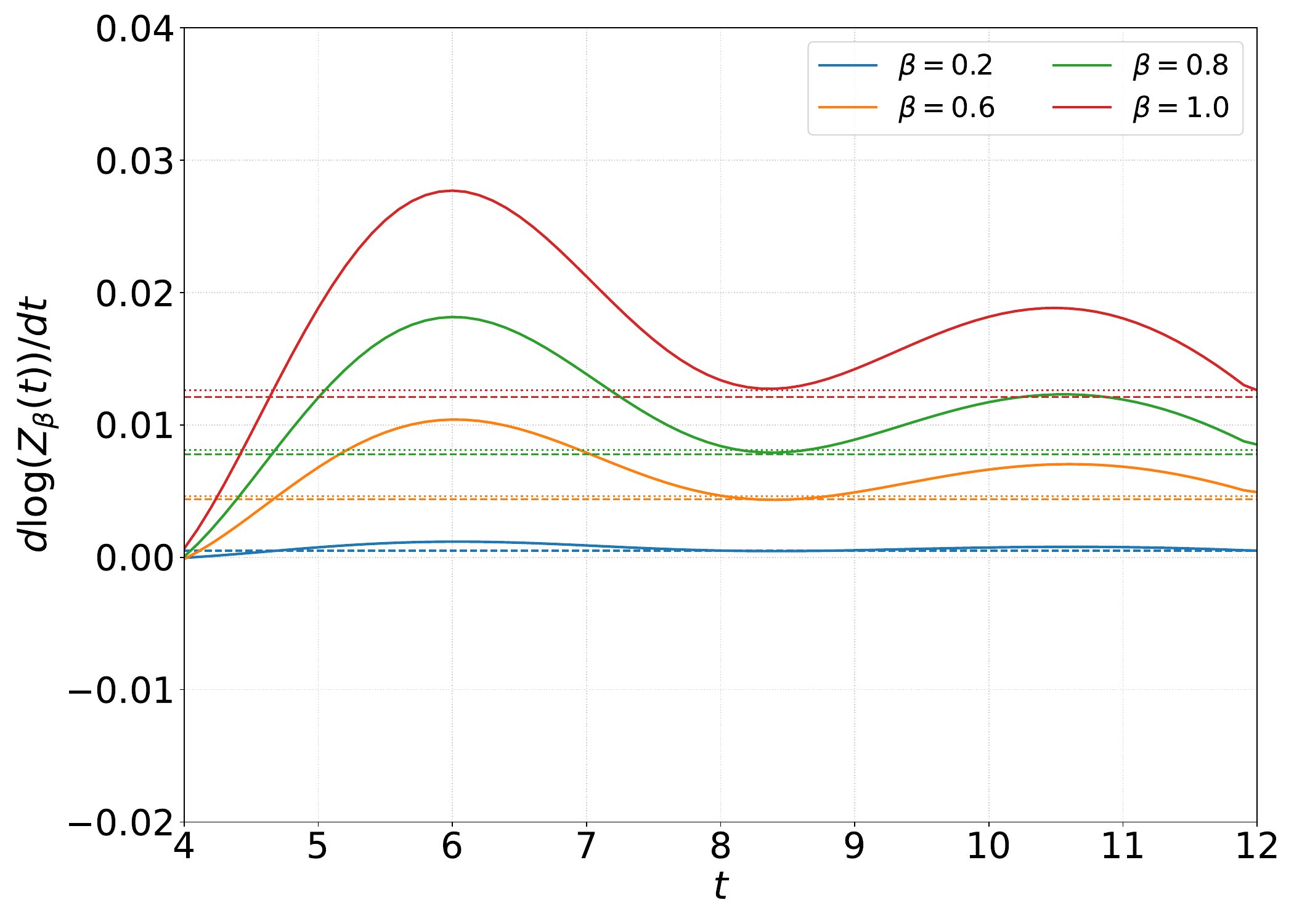}
    \end{subfigure}
    \hfill
    \begin{subfigure}[c]{0.48\textwidth}
        \centering
        \includegraphics[width=\linewidth]{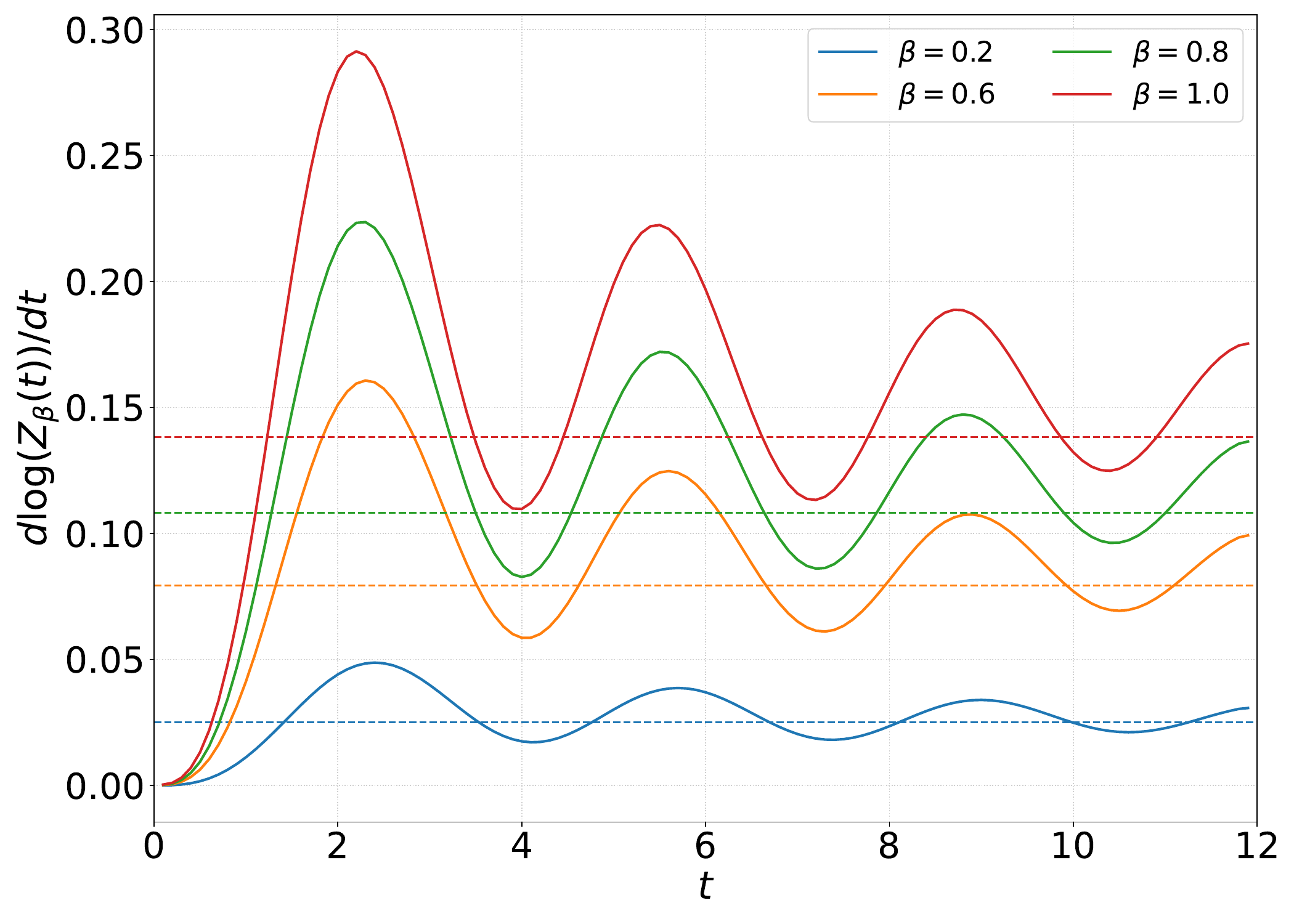}
    \end{subfigure}
    \caption{Dynamics of the FCS in the XXZ chain with $\Delta=0.5$ (gapless regime) for two different bipartite initial states: on the left a bipartition from two thermal states with inverse temperatures $\beta_L=5$ and $\beta_R=10$, and on the right a pure bipartition $\ket{F,0}\otimes\ket{N,0}$. Dashed lines represent the predictions obtained by using $\text{sgn}(\varepsilon'(\theta))$, while dotted lines are the predictions obtained using $\text{sgn}(v^{\rm dr}(\theta))$. It is clear that the FCS exhibits oscillations about the global slope which are much larger compared to the R\'enyi entropies, and this makes it impossible to distinguish the different predictions in this case. On the right side, a single predictions is present, since the sign of the dressed and bare velocities coincide.}
    \label{fig:fcs_quench_plot}
\end{figure}

We perform this analysis both for pure and mixed (thermal) bipartite states obtaining the results reported in Fig.~\ref{fig:fcs_quench_plot}, as well as Fig.~\ref{fig:comparisonneeldimer} of the main text. Although the results show convincing convergence to the analytical prediction in Eq.~\eqref{eq:FCSgeneralzeta}, the precision is not enough to distinguish between the latter and its variant  obtained by replacing ${\rm sgn}(\varepsilon'(\mu))$ with ${\rm sgn}(v^{\rm dr}(\mu))$. This happens both because the difference between the two predictions is smaller compared to the one observed for the entropy, and because the FCS exhibits stronger oscillations which do not disappear completely in the time interval accessible through TEBD. 

\begin{figure}[h!]
    % Row 2
    \begin{subfigure}[c]{0.48\textwidth}
        \centering
        \includegraphics[width=\linewidth]{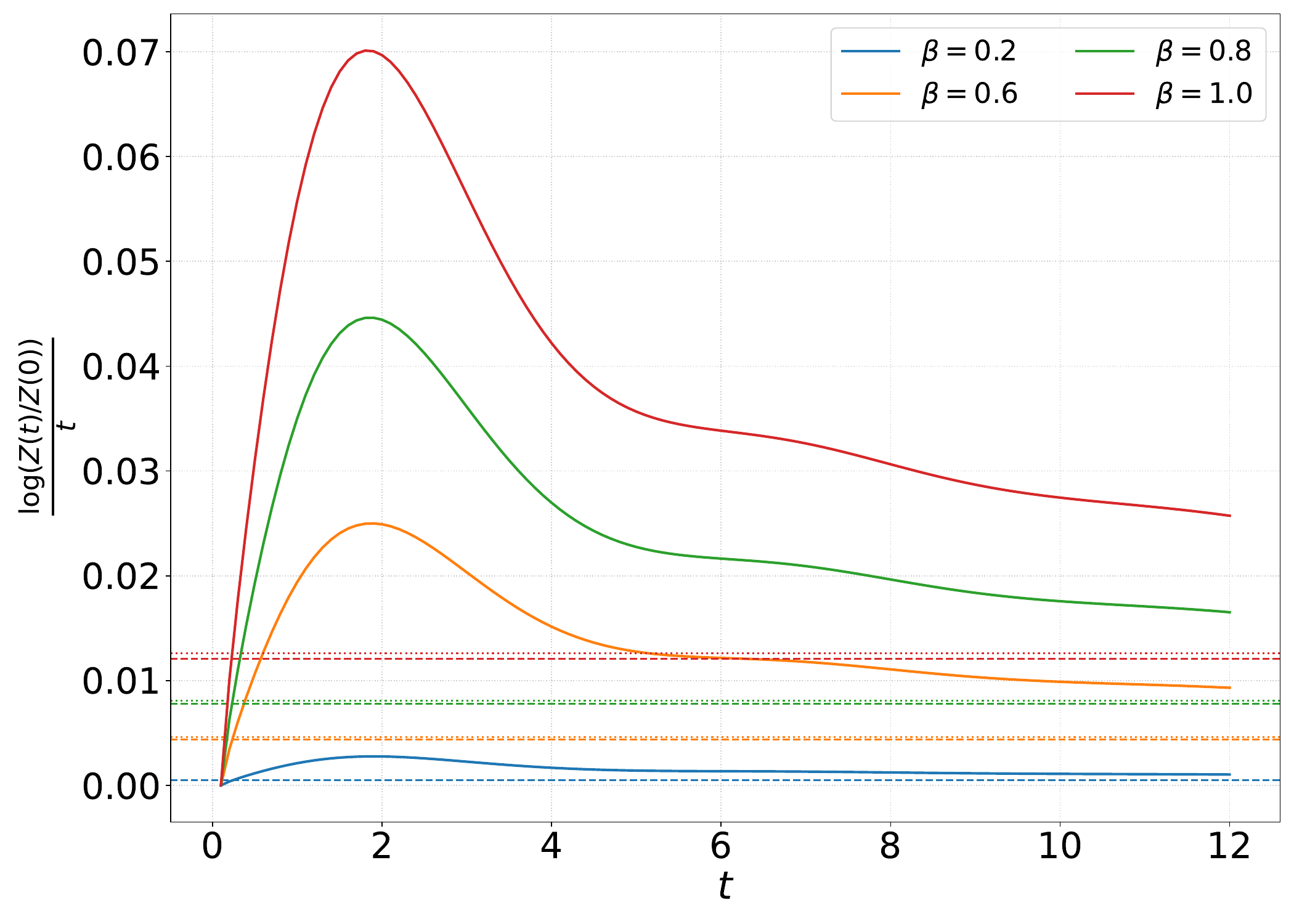}
    \end{subfigure}
    \hfill
    \begin{subfigure}[c]{0.48\textwidth}
        \centering
        \includegraphics[width=\linewidth]{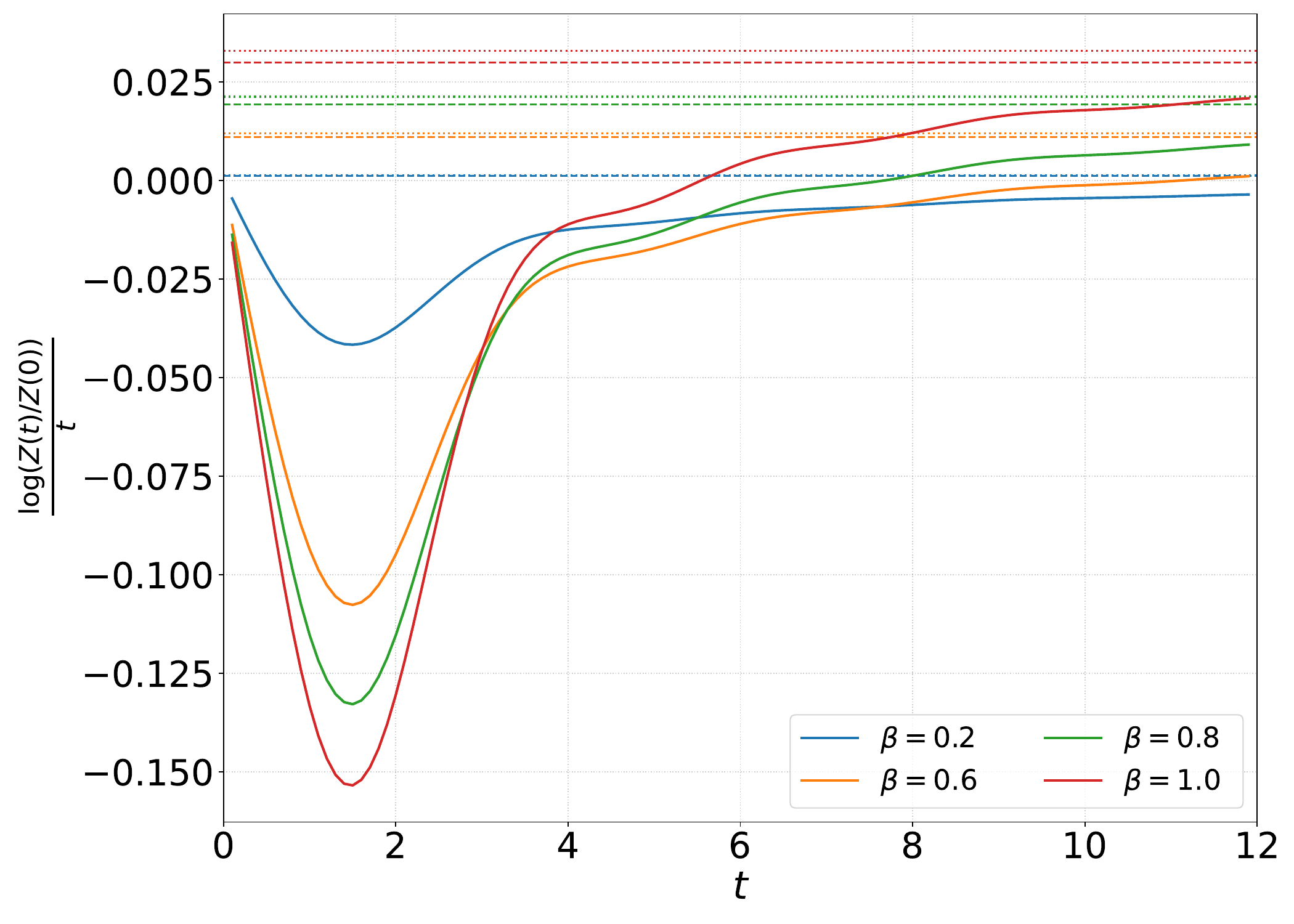}
    \end{subfigure}
    \caption{Dynamics of the FCS in the XXZ chain with $\Delta=0.5$ (gapless regime) for a quench from the thermal state of Fig. \ref{fig:fcs_quench_plot} (left), and the state considered in Fig.~\ref{fig:comparisonneeldimer} of the main text (right). Dashed lines represent the predictions obtained by using $\text{sgn}(\varepsilon'(\theta))$, while dotted lines are the predictions obtained using $\text{sgn}(v^{\rm dr}(\theta))$. ALthough evaluating the global slope allows to reduce the strong oscillations affecting the instantaneous one, the strong peak at early times makes it impossible to observe convergence to the asymptotic value in the accessible timescales.}
    \label{fig:fcs_quench_plot_global}
\end{figure}

The situation is not solved by focusing on the global slope, which was considered for example in \cite{bertini2023nonequilibrium,bertini2024dynamics}, as the strong fluctuations in the early-time cannot be canceled completely for the relatively small values of $t$ accessible in the simulations. This effect is clearly visible in Fig.~\ref{fig:fcs_quench_plot_global}. 

\end{document}